\documentclass[twocolumn,showpacs,preprintnumbers,amsmath,amssymb]{revtex4}
\usepackage{graphicx}
\usepackage{dcolumn}
\usepackage{bm}
\begin{document}
\preprint{APS/123-QED}
\title{Global stability analysis of axisymmetric boundary layer over a
       circular cone }
\author{Ramesh Bhoraniya}
\altaffiliation[Also working at]{ Dept. of Mechanical Engineering, Faculty of Engineering, 
Marwadi Education Foundations Groups of Institutions, Rajkot, INDIA.}
\author{Narayanan Vinod }%
\email{vinod@iitgn.ac.in}
\affiliation{%
Department of Mechanical Engineering, Indian Institute of Technology, Gandhinagar, INDIA\\
}%
%
\date{\today}
\begin{abstract}
This paper presents the linear Global stability analysis of the 
incompressible axisymmetric boundary layer on a circular cone.
The base flow is considered parallel to the axis of cone at the 
inlet.
The angle of attack is zero and hence the base flow is axisymmetric. 
The favorable  pressure gradient develops in the stream-wise direction 
due to cone angle.
The Reynolds number is calculated based on the cone radius (a) at the
inlet and free-stream velocity ($U_{\infty}$). 
The base flow velocity profile is fully non-parallel and non-similar.
Linearized Navier-Stokes equations (LNS) are derived for the disturbance 
flow quantities in the spherical coordinates.
The LNS are discretized using Chebyshev spectral collocation method.
The discretized LNS along with the homogeneous boundary conditions forms 
a general eigenvalues problem.
Arnoldi's iterative algorithm is used for the numerical solution of 
the general eigenvalues problem.
The Global temporal modes are computed for the range of 
Reynolds number from 174 to 1046, semi-cone angles $2^o$, $4^o$, $6^o$  
and azimuthal wave numbers from 0 to 5.
It is found that the Global modes are more stable at higher 
semi-cone angle $\alpha$, due to the development of  favorable  
pressure gradient.
The effect of transverse curvature is reduced at higher semi-cone 
angles ($\alpha$).
The spatial structure of the eigenmodes show that the flow is 
convectively unstable.
The spatial growth rate ($A_x$) increases with the increase in semi-cone 
angle ($\alpha$) from $2^o$ to $6^o$.
Thus, the effect of increase in semi-cone angle ($\alpha$) is to reduces 
the temporal growth rate ($\omega_i$) and to increases the spatial growth 
rate ($A_x$) of the Global modes at a given Reynolds number.
\end{abstract}
\pacs{Valid PACS appear here}
\maketitle
\section{\label{sec:level1}Introduction}
The laminar-turbulent transition in  boundary layers
has been a subject of interest to many researchers in past 
few decades.
It is important to understand the onset of transition 
in boundary layers as the  flow pattern and its effects are
very different in laminar  and turbulent flows.
For example, the drag force in a turbulent flows
is much higher than that of a laminar flows.
At low free-stream levels the boundary layers undergo transition through the
classical TS wave mechanism. 
The amplification of the disturbance waves is the primary step in the transition 
process and this is studied in linear stability analysis.
The results from stability analysis and the prediction of transition onset is very 
useful in hydrodynamics and aerodynamics applications like submarines,
torpedoes, rockets, missiles etc.

The linear stability analysis of shear flows with parallel flow assumption is well understood
by the solution of the Orr-Sommerfeld equation \cite{Drazin}. 
It is known that the stability characteristics in a boundary layer
is strongly influenced by various factors such as  pressure gradient, surface curvature
free-stream turbulence level. 

\begin{figure}
\begin{center}
\includegraphics [height=1.25in,width=1.65in, angle=0]
                 {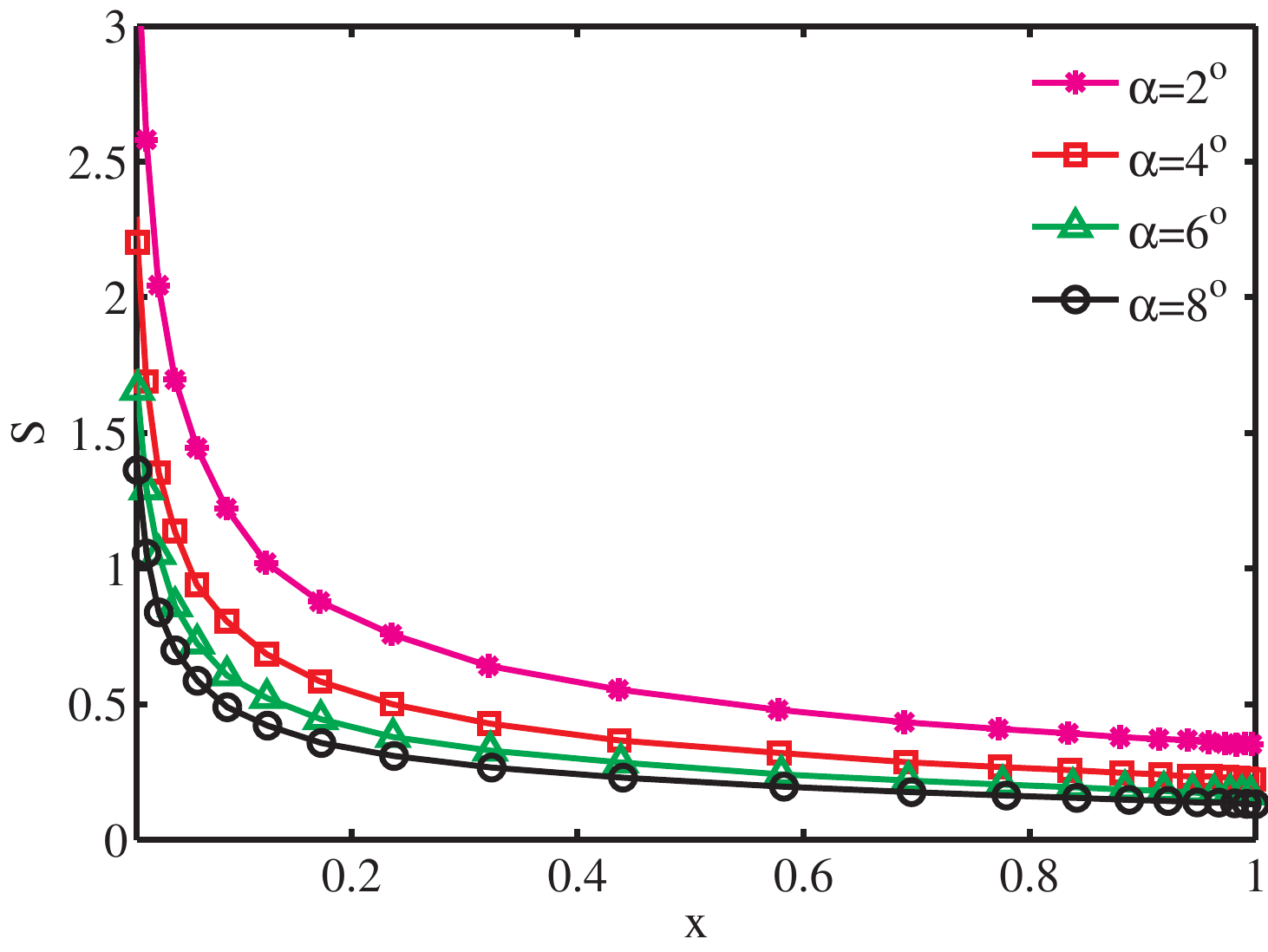}  
\includegraphics [height=1.25in,width=1.65in, angle=0]
                 {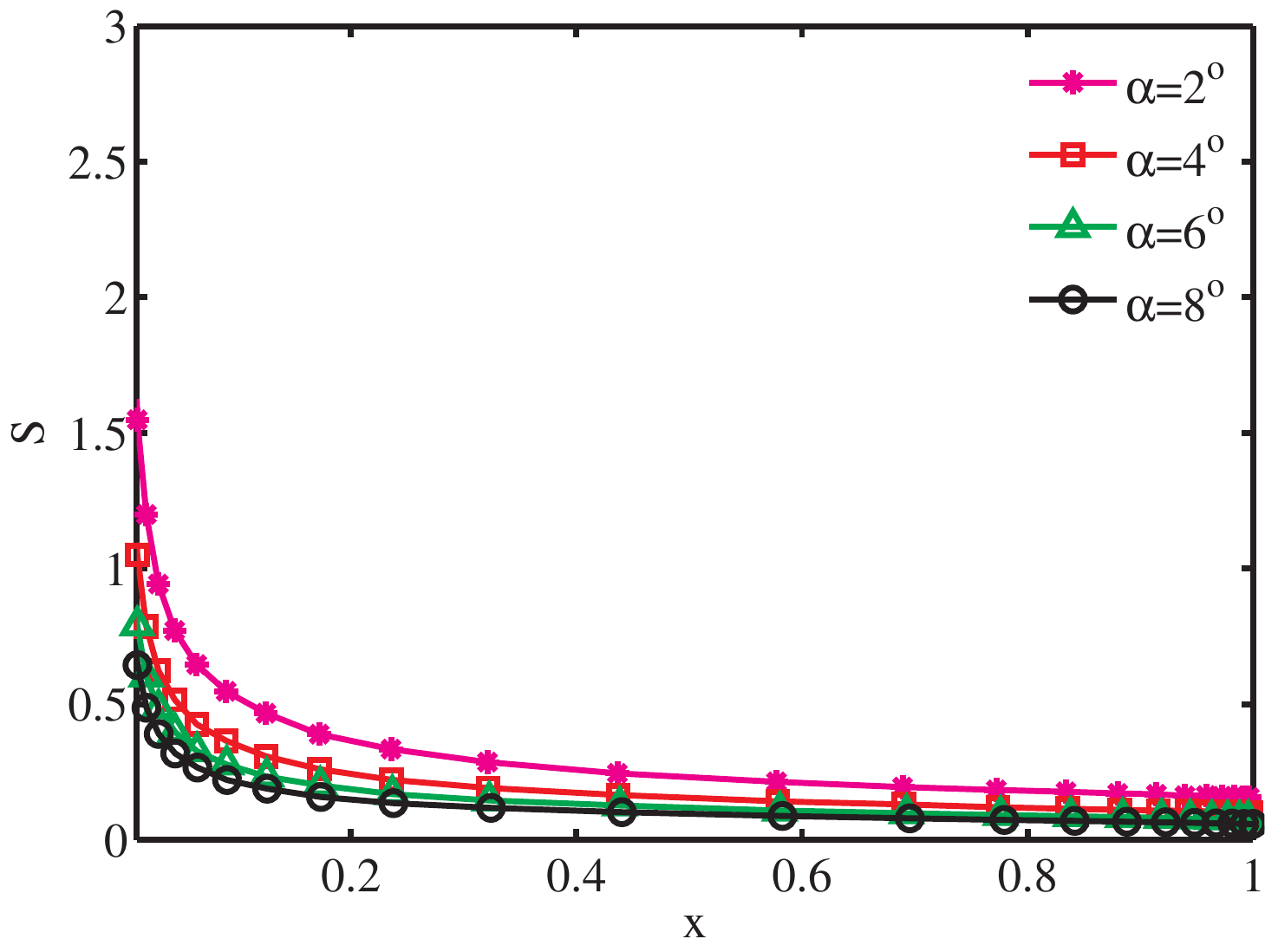} 
\caption{\label{curvature} Variation of the transverse curvature($S=\delta/a$)
         in the stream-wise direction for different semi-cone angles($\alpha$)
         for (a) Re=174 and (b) Re=872.}
\end{center}
\end{figure}

The boundary layer forms over a right circular cone is axisymmetric
and it is qualitatively different from a flat-plate boundary layer.
The boundary layer on a circular cone develops continuously in spatial directions, hence 
the parallel flow assumption is not valid. Due to the wall normal velocity
component, the boundary layer is non-parallel.
The instability analysis of such two-dimensional base flow is called Global stability 
analysis \cite{Theofilies03}.
The literature on the instability analysis of axisymmetric boundary layer is very sparse.
The available literature on the axisymmetric boundary layer is limited to local stability analysis. 
Rao \cite{Rao} first studied the stability of axisymmetric boundary layer. 
He found that non-axisymmetric disturbances are less stable than that 
of two-dimensional disturbances. 
Tutty \cite {Tutty} investigated that for non-axisymmetric modes critical 
Reynolds number increases with the azimuthal wave number. 
The critical Reynolds number found to be 1060 for $N=1$ mode and 12439 for $N=0$ mode. 
The axisymmetric mode is found least stable fourth mode. 
Vinod \cite{vinodthesis} investigated that higher non-axisymmetric mode $N=2$ is 
linearly stable for a small range of curvature only. 
The helical mode N=1 is unstable over a significant length of the cylinder, 
but never unstable for curvature above 1. 
Transverse curvature has overall stabilizing effect over mean flow and perturbations. 
Malik \cite{Malik85} studied the effect of transverse curvature on the stability of 
incompressible boundary layer. 
They investigated that the body curvature and streamline curvature are having 
significant damping effects on disturbances. 
The secondary instability of an incompressible axisymmetric boundary 
layer is also studied by Vinod \cite{vinod12}.
They found that laminar flow is always stable at high transverse curvature 
to secondary disturbances.
The Global stability analysis for hypersonic and supersonic flow over a circular 
cone is reported in the literature. 
In supersonic and hypersonic boundary layers there exists higher acoustic instability 
modes with higher frequencies in addition to first instability mode \cite{Mack}.
It has been confirmed by the experiments that the two dimensional Mack mode dominates 
in hypersonic boundary layer \cite{stetson1,stetson2,Maslov}.
Experimental results show that leading bluntness affects the transition location significantly 
in circular cone \cite{Horvath}.
In case of axisymmetric boundary layer on a circular cylinder, transverse curvature 
effect increases in the stream-wise direction, which helps in stabilizing the flow.
In case of a axisymmetric  boundary layer on a circular cone, the body radius of the cone
increases at higher rate than a boundary layer thickness ($\delta$) and hence the 
transverse curvature effect ($\delta/a$) reduces in the stream-wise direction as shown 
in figure \ref{curvature}.
However, the favorable  pressure gradient develops in the stream-wise direction 
due to cone angle ($\alpha$).
Thus, it becomes interesting to study the combined effect of transverse curvature and
pressure gradient on the stability characteristics. 
The two dimensional global modes are also computed for the flat-plate boundary layer 
by some investigators \cite {Espen,Alizard,Ehrenstein}.
However, this is the first attempt to carry out Global stability
analysis of incompressible boundary layer over a circular cone.
The main aim of this paper is to study the Global stability characteristics 
of the axisymmetric boundary layer on a circular cone and the effect of the 
transverse curvature  and pressure gradient on the stability characteristics.   
\section{\label{sec:level1} Problem formulation }
The standard procedure is followed for the derivation of the Linearized
Navier-Stokes equations(LNS) for the disturbance flow quantities.
The Navier-Stokes (N-S) equations for the base flow and instantaneous flow 
are written in the spherical coordinate system (r,$\theta$,$\phi$).
The equations are non-dimensionalised using  free-stream velocity 
$ U_\infty$  and body radius of the cone $a$ at the inlet.
The LNS for disturbance flow quantities are obtained by subtraction 
and subsequent linearization.
The base flow is two dimensional and disturbances are three dimensional 
in nature.
This will determine whether the small amplitudes of the disturbances 
to grow or decay for a given steady laminar flow.
The Reynolds number is defined as,
\\
  \begin{equation}
     R_{e}= \frac{U_\infty a}{\nu}
  \end{equation}
where  $a$ is the  surface radius of cone at inlet 
              and $\nu$ is the kinematic viscosity. 

  The flow quantities are presented as sum of the base flow and 
  the perturb quantities as,
  \begin{equation}\\
     \overline {U_r} =U_r+u_r, \\ 
     \overline {U_\theta}=U_\theta+u_\theta ,\\ 
     \overline {U_\phi}=0+u_\phi, \\
     \overline {P}=P+p ,\\
  \end{equation}
   The disturbances are considered in normal mode form and varying in radial(r) and
   polar($\theta$) directions. Thus the disturbances having periodic nature in the 
   azimuthal ($\phi$) direction. 
  \begin{equation}\\ 
      q(r,\theta,t)= \hat q(r,\theta)e^{[i(N \phi-\omega t)]}
  \end{equation}
     where, \\
             $q=[u_{r},u_{\theta},u_{\phi},p]$\\
             $ Q =[U_r,U_{\theta},P]$ \\
          $\overline {Q}=[\overline U_r ,\overline U_{\theta},\overline P]$\\
              r = radial coordinate \\
              $\theta$ = polar coordinate \\
              $\phi$ = azimuthal coordinate \\
              $\omega$ = circular frequency \\
              $N$ = azimuthal wave-number \\
   \begin{figure}
   \centerline {\includegraphics [height=2.5in, width=3.5in, angle=0]
                                 {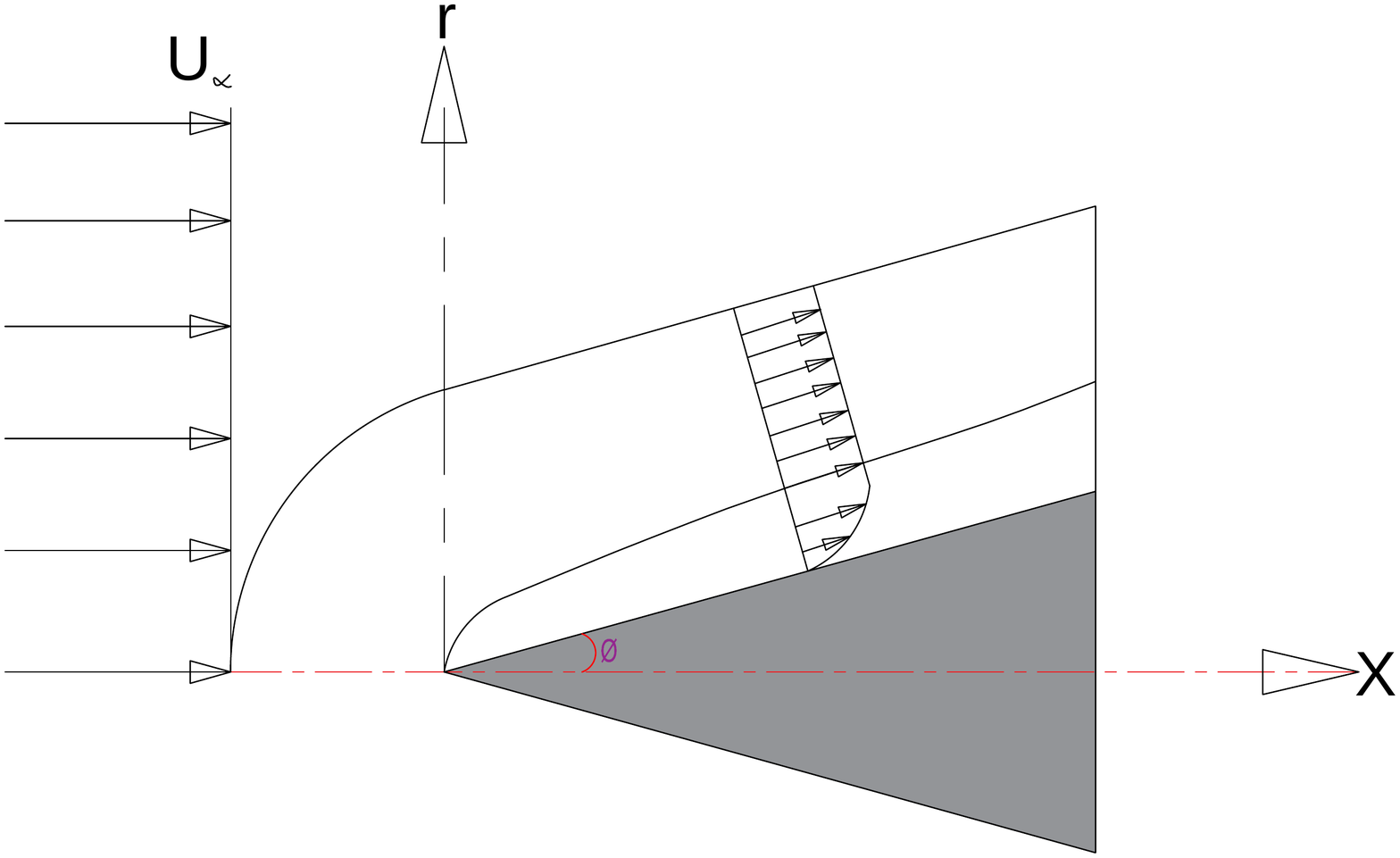}}  
   \caption{\label{aximodel} Schematic diagram of axisymmetric boundary 
                              layer on a circular cone.}
   \end{figure}
   \begin{multline}
         \frac{\partial u_r}{\partial t} + U_r \frac{\partial u_r}{\partial r} 
         + u_r \frac{\partial U_r}{\partial r} + \frac{U_\theta}{r} \frac{\partial u_r}
         {\partial \theta}+ \frac{u_\theta}{r} \frac{\partial U_r}{\partial \theta} 
         - u_\theta \frac{2 U_\theta}{r} + \frac{\partial p}{\partial r} 
         \\ -\frac {1}{Re} [\bigtriangledown^2 u_r - \frac{2 u_r}{r^2}-
         \frac{2}{r^2} \frac{\partial u_\theta}{\partial \theta} 
        -\frac{2 cot\theta}{r^2} u_\theta 
        - \frac {2} {r^2 sin\theta} \frac{\partial u_\phi}{\partial \phi}]=0
   \end{multline}
   \begin{multline}
       \frac{\partial u_\theta}{\partial t} + U_r \frac{\partial u_\theta}{\partial r}
       + u_r \frac{\partial U_\theta}{\partial r} + \frac{U_\theta}{r} 
       \frac{\partial u_\theta}{\partial \theta} + \frac{u_\theta}{r} 
       \frac{\partial U_\theta}{\partial \theta} + u_\theta \frac{U_\theta}{r}
       + u_r \frac{U_\theta}{r}\\ + \frac{1}{r}\frac{\partial p} {\partial \theta}
       -\frac {1}{Re} [\bigtriangledown^2 u_\theta
       +\frac{2}{r^2} \frac{\partial u_r} {\partial \theta} 
       -\frac{u_\theta}{r^2 sin^2\theta}-\frac{2 cot\theta} 
       {r^2 sin\theta} \frac{\partial u_\phi}{\partial \phi}]=0
     \end{multline}
    \begin{multline}
     \frac{\partial u_\phi}{\partial t} + U_r \frac{\partial u_\phi}{\partial r} 
     + \frac{U_\theta}{r} \frac{\partial u_\phi}{\partial \theta}
     + u_\phi \frac{U_r}{r}+ u_\phi \frac{U_r cot\theta}{r}  
     + \frac{1}{r sin\theta} \frac{\partial p}{\partial r} \\
     -\frac {1}{Re} [\bigtriangledown^2 u_\phi +\frac{2}{r^2 sin\theta} 
     \frac{\partial u_\phi}{\partial \phi} -\frac{ u_\phi}{r^2 sin\theta}
     +\frac {2 cot\theta} {r^2 sin\theta} \frac{\partial u_\phi}{\partial \phi}]=0
     \end{multline}
     \begin{equation}
      \frac{\partial u_r}{\partial r} + \frac{2u_r }{r} 
     + \frac{1}{r} \frac{\partial u_\theta}{\partial \theta} 
     + \frac{u_\theta cot\theta} {r}
     + \frac{1}{r sin\theta} \frac{\partial u_\phi}{\partial \phi}=0
     \end{equation}
      where,
     \begin{equation}
      \bigtriangledown^2 = \frac{\partial }{\partial r^2} 
      + \frac{2}{r} \frac{\partial}{\partial r}+ \frac{1}{r^2} 
      \frac{\partial}{\partial \theta^2} 
      +\frac{cot\theta}{r^2} \frac{\partial }{\partial \theta} 
      + \frac{1}{r^2 sin^2\theta} \frac{\partial }{\partial \phi^2}
     \end{equation}
     \subsection{\label{sec:level2} Boundary conditions}
      No slip and no penetration boundary conditions are considered at the 
      surface of the cone. 
      The magnitude of all the disturbance velocity components are zero 
      at the solid surface of the cone due to viscous effect.\\
     \begin{equation} 
      u_r(r,\theta_{min})=0 ,  u_\theta(r,\theta_{min})=0 , 
      u_\phi(r,\theta_{min})=0 
     \end{equation}  
     It is expected to vanish the disturbance flow quantities at free-stream
     far away from the solid surface of the cone.
     The Homogeneous Dirichlet conditions are applied to all 
     velocity and pressure disturbances at free-stream.
     \begin{multline}
       u_r(r,\theta_{max})=0, u_\theta(r,\theta_{max})=0, \\
       u_\phi(r,\theta_{max})=0, p(r,\theta_{max})=0
     \end {multline} 
      The boundary conditions are not straight forward in the stream-wise direction
      for Global stability analysis. 
      As suggested by Theofilis \cite{Theofilis03}, homogeneous Dirichlet boundary conditions
      are considered for velocity disturbances at inlet. 
      Here we are interested in the disturbances evolved within the basic flow 
      field only.
      The boundary conditions at outlet can be applied based on the incoming and outgoing
      wave information \cite{Fasel}.      
      Such conditions impose wave like nature of the disturbances and so it is 
      more restrictive in nature which is not appropriate from the physical 
      point of view.
      Even stream-wise wave number $\alpha$ is not known initially in case of the 
      Global stability analysis.  
      Alternatively  one can impose such numerical boundary condition which 
      extrapolate the information from the interior of the computational domain.
      Linear extrapolated conditions are applied by several investigators in such case.
      The review of the literature on Global stability analysis suggests that the  
      linear extrapolated boundary conditions are least restrictive in 
      nature \cite{Swaminathan,Theofilis03}.
      Thus, we considered linear extrapolated conditions at the outlet for 
      the numerical solution of the general eigenvalues problem.
      \begin{equation}
       u_r(r_{in},\theta)=0 ,  u_\theta(r_{in},\theta)=0 , 
       u_\phi(r_{in},\theta)=0 
      \end{equation}
      \begin{equation}
           u_{r_{n-2}}[r_n-r_{n-1}]-u_{r_{n-1}}[r_n-r_{n-2}] 
           + u_{r_{n}}[r_{n-1}-r_{n-2}]=0 
      \end{equation}
       similarly, one can write extrapolated boundary conditions for polar 
       and azimuthal disturbance components $u_{\theta}$ and $u_{\phi}$ respectively.
       The boundary conditions for pressure  do not exist physically at the wall.
       However compatibility conditions derived from the LNS equations are collocated 
       at the solid wall of the cone \cite{Theofilis03}.
      \begin{equation}\\
         \frac {\partial p}{\partial r}= \frac{1}{Re}[\bigtriangledown^2 u_r 
        - \frac{2}{r^2} \frac{\partial u_\theta}{\partial \theta}]
        - U_r\frac{\partial u_r} {\partial r} 
        - \frac{U_\theta} {r} \frac{\partial u_r}{\partial \theta}  
     \end{equation}
     \begin{equation}\\
        \frac{1}{r} \frac {\partial p}{\partial \theta}=\frac{1}{Re}
        [\bigtriangledown^2 u_\theta +\frac{2}{r^2} \frac{\partial u_r}
        {\partial \theta}] -U_r\frac{\partial u_\theta} {\partial r}
       - \frac{U_\theta} {r} \frac{ \partial u_\theta}{\partial \theta}  
      \end{equation}
       The Linearized Navier-Stokes equations are discretized using 
       Chebyshev spectral collocation method.
       The Chebyshev polynomial generates non uniform grids.
       By nature  it takes more number of points towards the ends.
       This is favorable  arrangement for the boundary value problems.
       \begin{equation}
        r_{cheb}= \cos(\frac{ \pi i}{n})\quad {\rm where} 
                  \quad i= 0,1,2,3...n
       \end{equation}
       \begin{equation}
        \theta_{cheb}= \cos(\frac{ \pi j}{m})\quad {\rm where} 
                       \quad j= 0,1,2,3...m
       \end{equation}
        The gradient of the velocity disturbances are very high near the solid 
        surface of the cone.
        To improve the spatial resolution in the boundary  layer region near the wall, 
        it is necessary to stretch more number of collocation points near the wall.
        The following algebraic equation is used for grid stretching (\cite{Malik}).
       \begin{equation}
         \theta_{real}= \frac{\theta_i*L_\theta*(1-\theta_{cheb})}
                        {L_\theta+\theta_{cheb}*(L_\theta-2 \theta_i)}
       \end{equation}
        In the above grid stretching  method half number of the collocation points 
        are concentrated within the $\theta_i$ angle from the lower surface 
        $\theta_{min}$ only.
        The non uniform nature of the distribution for the collocation points 
        in the stream-wise direction is undesirable.
        The maximum and minimum distance between the grid points are at the center 
        and at the end respectively in stream-wise direction.
        Thus it makes poor resolution at the center of the domain and very small 
        distance between the grids at the end gives rise to Gibbs 
        oscillations. To improve the spatial resolution and to minimize  the Gibbs 
        oscillation in the solution, grid mapping is implemented in   
        stream-wise direction using following algebraic equation \cite{Costa}.
        \begin{equation}
         r_{map}= \frac {sin^{-1}(\alpha_m r_{cheb})}{sin^{-1}(\alpha_m)}
        \end{equation}
         The value of $\alpha_m$ is chosen  carefully to improve the spatial resolution
         in the stream-wise direction.
         Very small value of $\alpha_m$ keeps the grid distribution almost like
         a Chebyshev and near to unity almost uniform grid distribution.
         For detail description of grid mapping readers are suggested to refer \cite{Costa}.
         To incorporate the effect of physical dimensions of the domain [$L_r$, $L_\theta$] 
         along with  grid stretching and mapping it is required to multiply the  Chebyshev 
         differentiation matrices by proper Jacobean matrix.
         Once all the partial derivatives of the LNS are discretized by spectral collocation 
         method using Chebyshev polynomials, the operator of the differential equations generate 
         matrices A and B. These matrices are square, real and sparse in nature, 
         formulates general eigenvalues problem.   
        $$
        \left[\begin{array}{cccc} A_{11} & A_{12} & A_{13} & A_{14} \\ 
             A_{21} & A_{22} & A_{23} & A_{24} \\
             A_{31} & A_{32} & A_{33} & A_{34} \\ 
             A_{41} & A_{42} & A_{43} & A_{44}  
             \end{array} \right] 
       \left[ \begin{array}{c} u_r\\ u_\theta \\ u_\phi \\ p \end{array} 
       \right]
        $$

        \begin{equation}\\
        =i\omega \left[ \begin{array}{cccc} 
            B_{11} & B_{12} & B_{13} & B_{14} \\ 
            B_{21} & B_{22} & B_{23} & B_{24} \\ 
            B_{31} & B_{32} & B_{33} & B_{34} \\
            B_{41} & B_{42} & B_{43} & B_{44} \end{array}  
       \right] \left[\begin{array}{c} u_r \\u_\theta\\u_\phi \\ p \end{array}
       \right]
        \end{equation}      
         \begin{equation}
          [A][\phi]=i\omega[B][\phi]\\
         \end{equation}
         where $A$ and $B$ are square matrix of size $4nm$, $ i\omega$ is an eigenvalues 
         and $\phi$ is a vector of unknown disturbance amplitude functions $ u_r $,$u_\theta$,
         $u_\phi$ and $p$.
         The above mentioned all the boundary conditions 
         are  properly incorporated in the matrix A and B.
         \subsection{\label{sec:level2} Solution of general eigenvalues problem}
         The A and B matrix are very large in size and sparse in nature.
         The full solution of the general eigenvalue problem is very difficult due to
         large size of matrix A and B.
         However, the shear flow becomes unstable due to very few leading eigenmodes only.
         The eigenmodes with largest imaginary part are important for instability analysis.
         Hence we are interested  in the few eigenvalues and associated eigen vectors only.
         The QZ algorithm is not the right choice for the solution, because it computes 
         full spectrum.
         Arnoldi's iterative algorithm is selected to compute the few selected eigenvalues.
         The Krylov subspace provides possibility of extracting major part of the spectrum 
         using  shift and invert strategy.
         The computations of Krylov subspace along with Arnoldi's algorithm applied to
         eigenvalues problem becomes easy.  
         \begin{equation}
          (A-\lambda B)^{-1} B\phi= \mu\phi \quad  
             where \quad \mu=\frac{1}{i\omega-\lambda}
         \end{equation}
         where, $ \lambda$ being the shift parameter and $\mu$ is the 
         eigenvalues of the converted problem.
         Sometimes it is also called spectral transformation, which converts 
         generalized eigenvalues problem to standard eigenvalues problem.
         The Krylov subspace may be computed by successive resolution of linear 
         system with matrix $(A-\lambda B)$,using LU  decomposition.
         Full spectrum method is employed for this small subspace to get good 
         approximate solution of the original general eigenvalues problem \cite{Theofilis11}.
         The major part of the spectrum can be extracted with large size of Krylov subspace.
         Generally the computed spectrum is always nearby the shift parameter.
         The good approximation of shift parameter reduces the number of iterations 
         to converge the solution to required accuracy level.
         However larger size of subspace  makes the solution almost independent from the 
         shift parameter $\lambda$.
         We tested the code for several values of shift parameter. 
         The convergence of the solution depends on the value of shift parameter.
         Good approximation of the shift value needs less number of iterations.
         However we have taken maximum number of iteration equal to 300, hence 
         convergence of the solution may not be affected by shift parameter.
         Given the large subspace size of $k=250$,the part of spectrum 
         for our instability analysis could be recovered in the one computation 
         which took about 4 hours on Intel Xenon(R)CPU E5 26500@$2.00GHz \times 18$.
         \section{\label{sec:level1} Base flow solution }
          The base flow velocity profile is obtained by numerical solution of the
          steady incompressible axisymmetric Navier-Stokes equations using 
          finite-volume code ANSYS FLUENT.
          The incoming flow is parallel to the axis of the cone at inlet.
          The axisymmetric domain is modeled with 1m and 0.5m in axial 
          and radial directions respectively as shown in figure \ref{aximodel}.
         \begin{equation}
              U \frac{\partial U} {\partial x}+V\frac{\partial U} {\partial r}= 
              - \frac{\partial P}{\partial x} +\frac{1}{R_{e}} (\frac{\partial^2 U}{\partial x^2} 
              + \frac{1}{r}\frac{\partial U}{\partial r}+ \frac{\partial^2 U}{\partial r^2})\\
         \end{equation}
         \begin{equation}
             U \frac{ \partial V} {\partial x} + V \frac{\partial V} {\partial r}= 
             - \frac{\partial P}{\partial r}+\frac{1}{R_{e}} (\frac{\partial^2 V}{ \partial x^2} 
              + \frac{1}{r} \frac{\partial V}{\partial r}+ \frac{\partial^2 V}{\partial r^2})\\
         \end{equation}
         \begin{equation}
             \frac{\partial U}{\partial x}+ \frac{\partial V}{\partial r}+ \frac{V}{r}=0
         \end{equation}
 \begin{table}
 \caption{\label{meangci}
 The grid convergence study for the base flow is obtained, 
 using U (x=0.5, r=0.038965 ) and V ( x=0.5, r=0.038965) for 
 $U_\infty$=0.1 m/s. 
 The grid refinement ratio ($\alpha$)in each direction is 1.4142.
 The relative error ($\epsilon$) and Grid Convergence Index (GCI) 
 are calculated using two consecutive grid size. 
 The j and j+1 represents course and fine grids respectively.
 $\epsilon$=$\frac{f_{j}-f_{j+1}} {f_{j}}$ $\times$ 100. 
 $GCI(\%)$=$3[\frac{f_j-f_{j+1}}{f_{j+1}(\alpha^n-1)}]$ $\times$ 100, 
 where n=log[$\frac{f_{j}-f_{J+1}}{f_{j+1}-f_{j+2}}$]/log({$\alpha$}).}   
\begin{ruledtabular}
\begin{tabular}{ccccccccc}
 $Mesh$&$U$& $\epsilon(\%)$&$GCI(\%)$ 
 & $V$&$\epsilon(\%)$&$GCI(\%)$ \\
\hline
  \# 1  & 0.087225 & 0.008
& 0.038 & 0.005893 & 0.017&0.051 \\
  \# 2  & 0.087213 & 0.0287
& 0.080 & 0.005892 & 0.034&0.1028 \\
  \# 3  & 0.087188 & 0.0510
& 0.144 & 0.005890 & 0.068&0.2037 \\
  \# 4  & 0.087143 & ---
& ---  & 0.005886 & --- & --- \\
\end{tabular}
\end{ruledtabular}
\footnotetext {Where, $\# 1= 707 \times 355$, $\# 2= 500 \times 250$,
                $\# 3=355 \times 177 $, \\ $\# 4=250 \times 125 $ }
\end{table}
\begin{figure}
\begin{center}
\includegraphics [height=1.20in,width=2.50in, angle=0]
                 {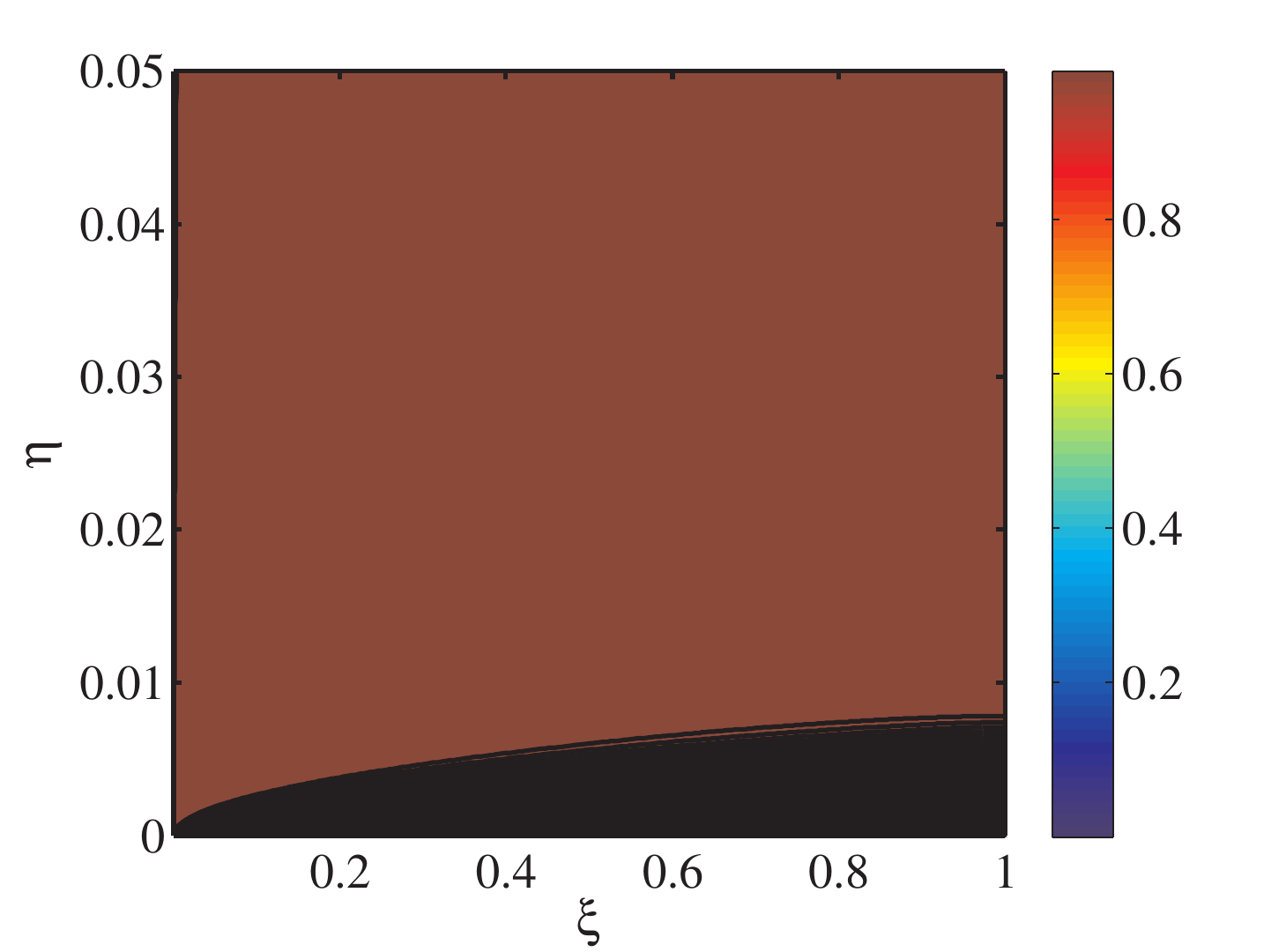}  
\includegraphics [height=1.20in,width=2.50in, angle=0]
                 {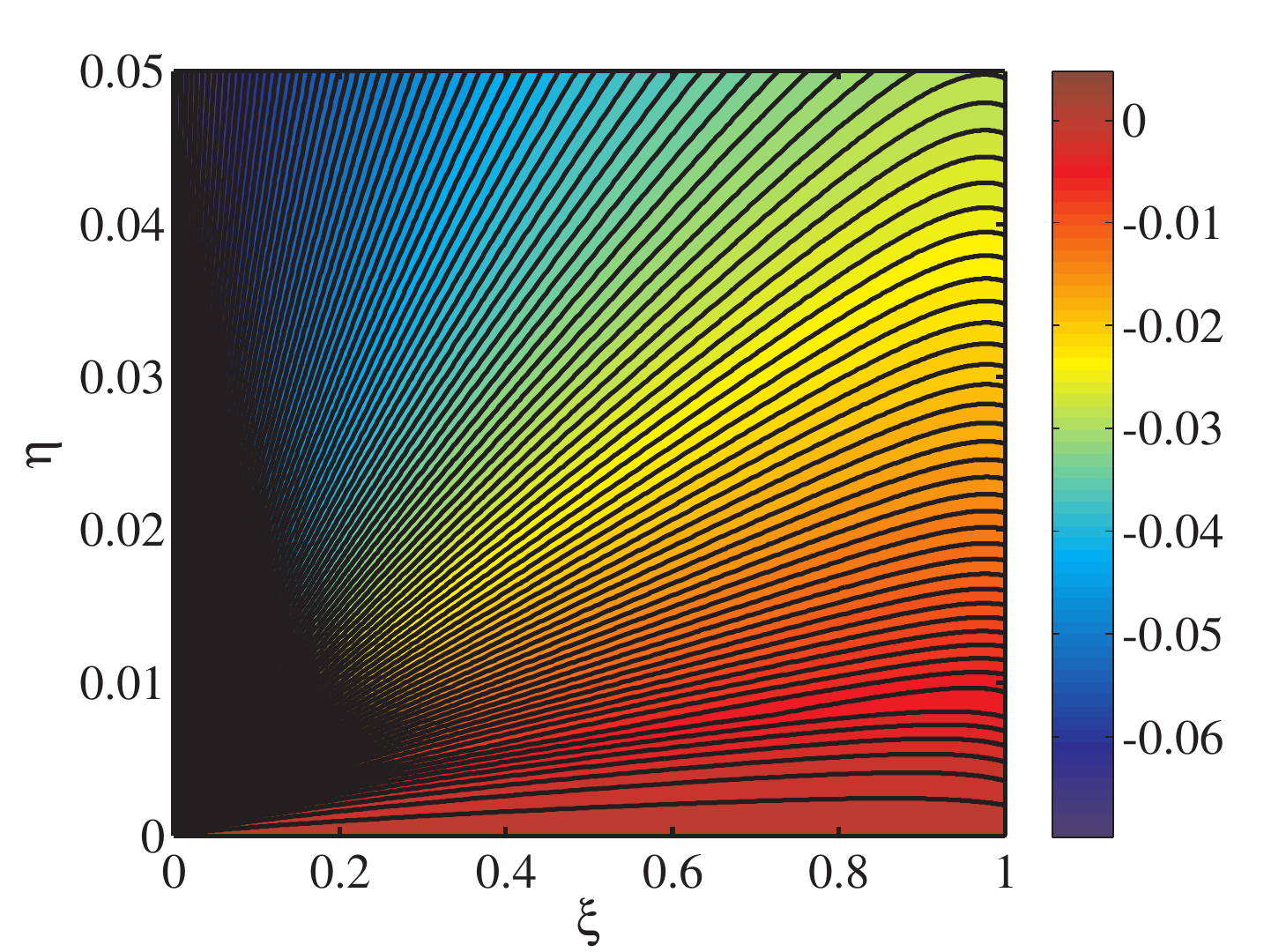} 
\caption{\label{contbaseflow} Base flow contour plot for (a) stream-wise $U_{\xi}$ 
         and (b) wall normal $U_{\eta}$ velocity components for semi-cone angle 
         $\alpha=4^o$ and Re=698. The Reynolds number is calculated based on 
         the body radius of the cone (a) at inlet of the domain. The velocity
         profile is interpolated to spectral grids for stability analysis. 
         The actual domain height is taken 0.5m in wall normal direction for 
         base flow computations. Here $\xi$ and $\eta$ are the coordinates in
         stream-wise and wall normal directions respectively.}
\end{center}
\end{figure}
\begin{figure}
\begin{center}
\includegraphics [height=1.25in,width=1.65in, angle=0]
                 {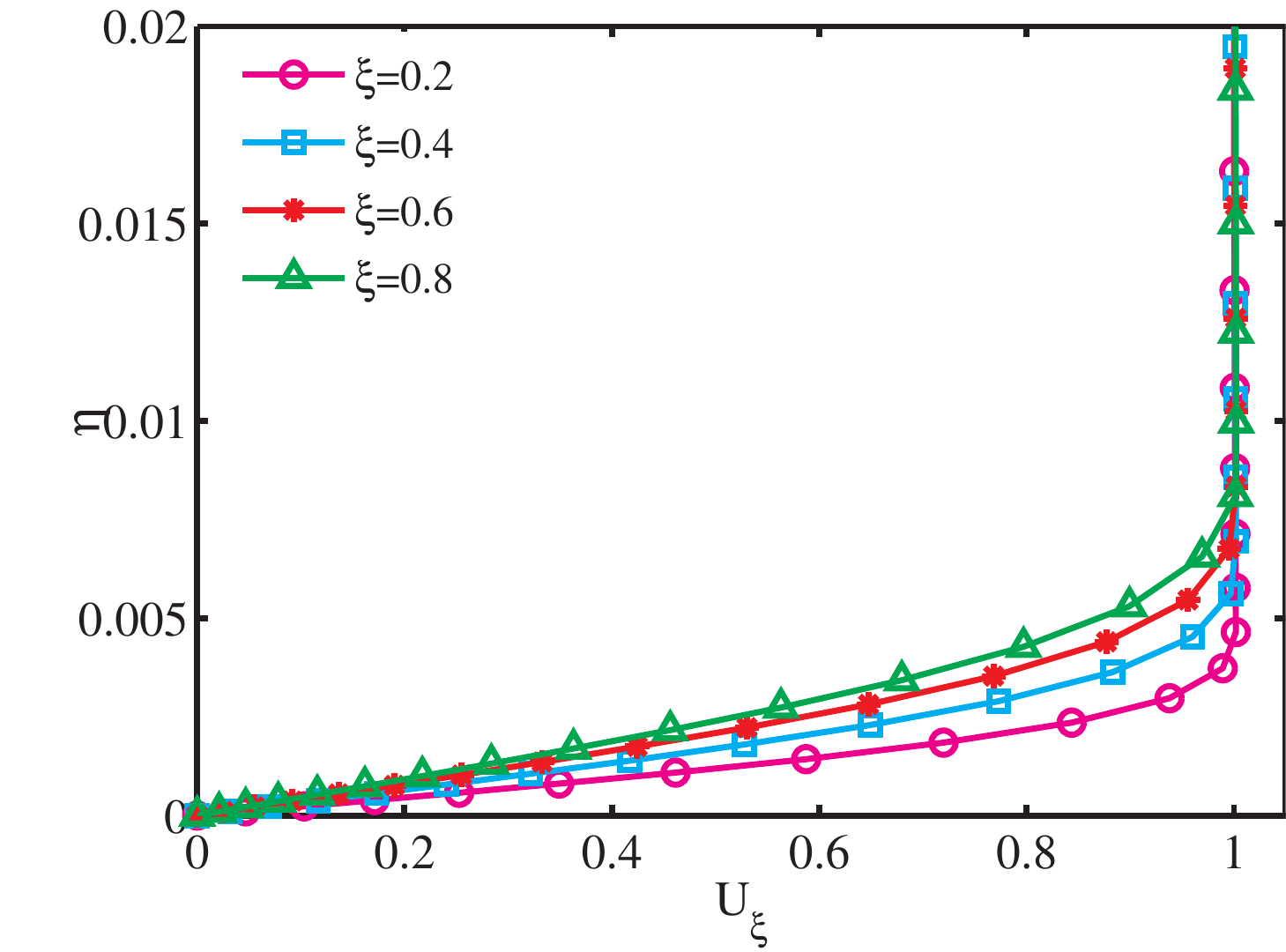}  
\includegraphics [height=1.25in,width=1.65in, angle=0]
                 {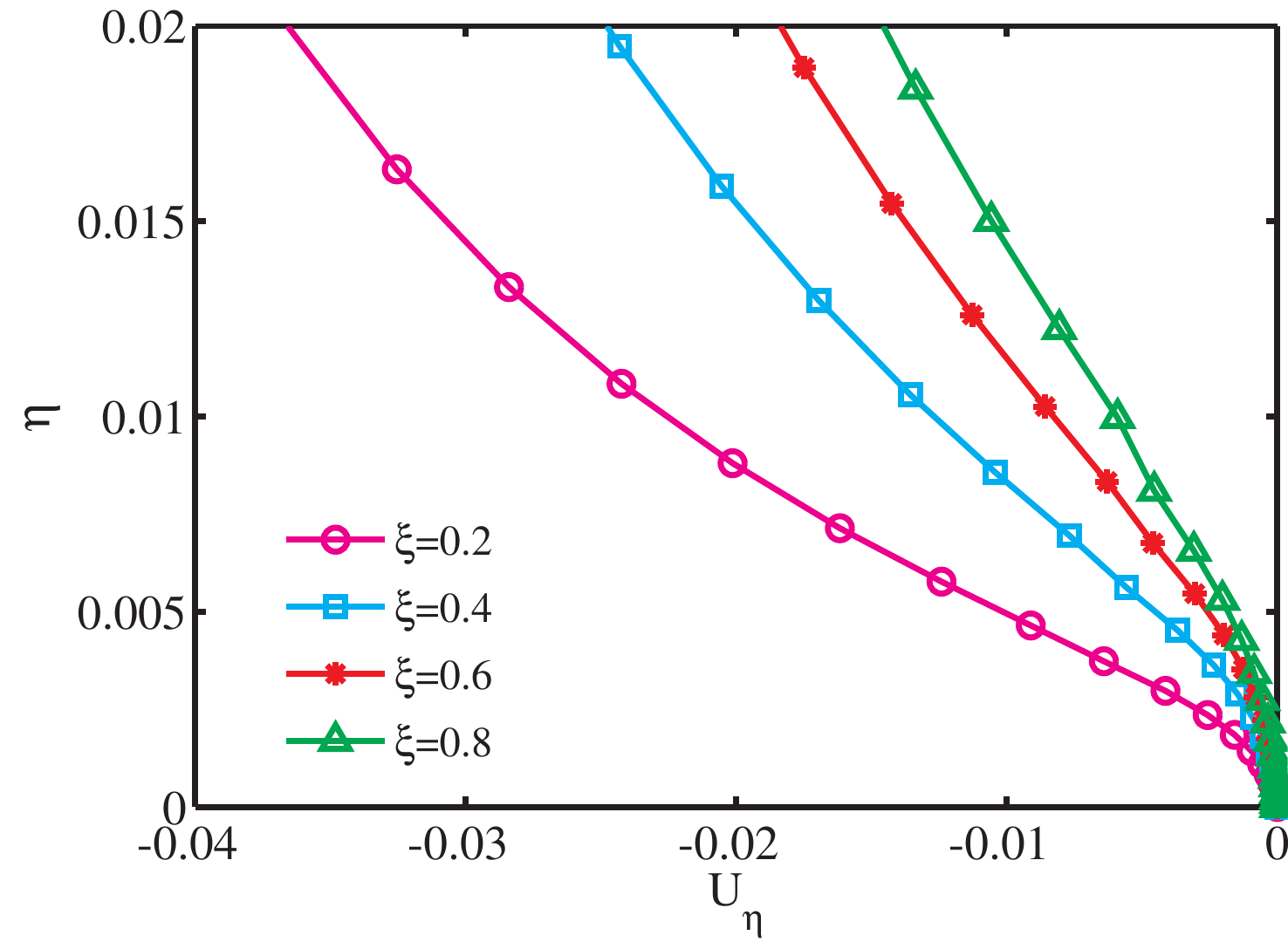} 
\caption{\label{vpbaseflow} Base flow velocity profile at different 
         stream-wise locations for (a)stream-wise velocity $U_{\xi}$ and
         (b)wall normal velocity $U_{\eta}$ for same as shown in figure
         \ref{contbaseflow}.
}
\end{center}
\end{figure}
Appropriate boundary conditions are applied to close the formulation 
of the above problem.
The uniform inlet velocity $U_{\infty}$ is imposed at the inflow boundary.
No-slip and no penetration conditions are applied on the surface of the cone.
At far away from the solid surface, slip boundary condition is applied.
The outflow boundary conditions are applied at outlet, which consists 
$ \frac{\partial U}{\partial x}=0 $, $\frac{\partial V}{\partial x}=0$ 
and $ P=0 $. 
The steady Navier-Stokes equations are solved using SIMPLE 
algorithm with under-relaxation to get the stable solution.
The spatial discretization of the N-S equations are done using a Quick 
scheme, which is a weighted  average of second order upwind and second order 
central scheme.
Thus obtained base velocity profile is interpolated to a spectral grids 
using cubic spline interpolation.
The simulation was started initially with the coarse grid size of 250 and 125
grid points in axial and radial directions respectively.
The grid was refined with a factor of 1.4142 in each directions.
The discretization error was calculated through Grid Convergence Index (GCI)
study on four consecutive refined grids \cite{Roache}.
The monotonic convergence is obtained for all this grids as expected.
The error and GCI are computed for two different field values U (x=0.5, r=0.038965) 
and V (x=0.5, r=0.038965) near the solid boundary where the gradient is higher.
The GCI and error were computed for four different grids as shown in 
table \ref{meangci}.
The error between Mesh 1 and Mesh 2 is too small.
The GCI has also reduced with the refinement of the grids.
Thus,the solution has converged monotonically towards the grid independent one.
The further refinement in the grids will hardly improve the accuracy of the 
solution while increases the time for the computations.
The distribution of grid is geometric in both the directions.
The computed convergence order for U and V are 1.98 and 2.01 are in 
good agreement with the second order discretization scheme used in the finite 
volume code ANSYS FLUENT.
Sufficient large domain height is selected in radial direction, i.e 0.5m.
Thus the Mesh 1 is used in all the results presented in to calculate 
velocity profile for basic state. 
Figure \ref{contbaseflow} and \ref{vpbaseflow} show the contour plot and velocity profile 
on the transformed coordinates $\xi$ and $\eta$.
The velocity profile is qualitatively similar to that of Garratt(2006) \cite{Garrett2}. 
The governing equations for instability analysis are derived in spherical 
coordinates.
The base velocity field is transformed  to spherical coordinates to perform 
Global stability analysis. 
\section{\label{sec:level1} Code validation}
\begin{figure}
\begin{center}
\vspace{12pt}
\includegraphics [height=3.0in,width=3.50in, angle=0]
                 {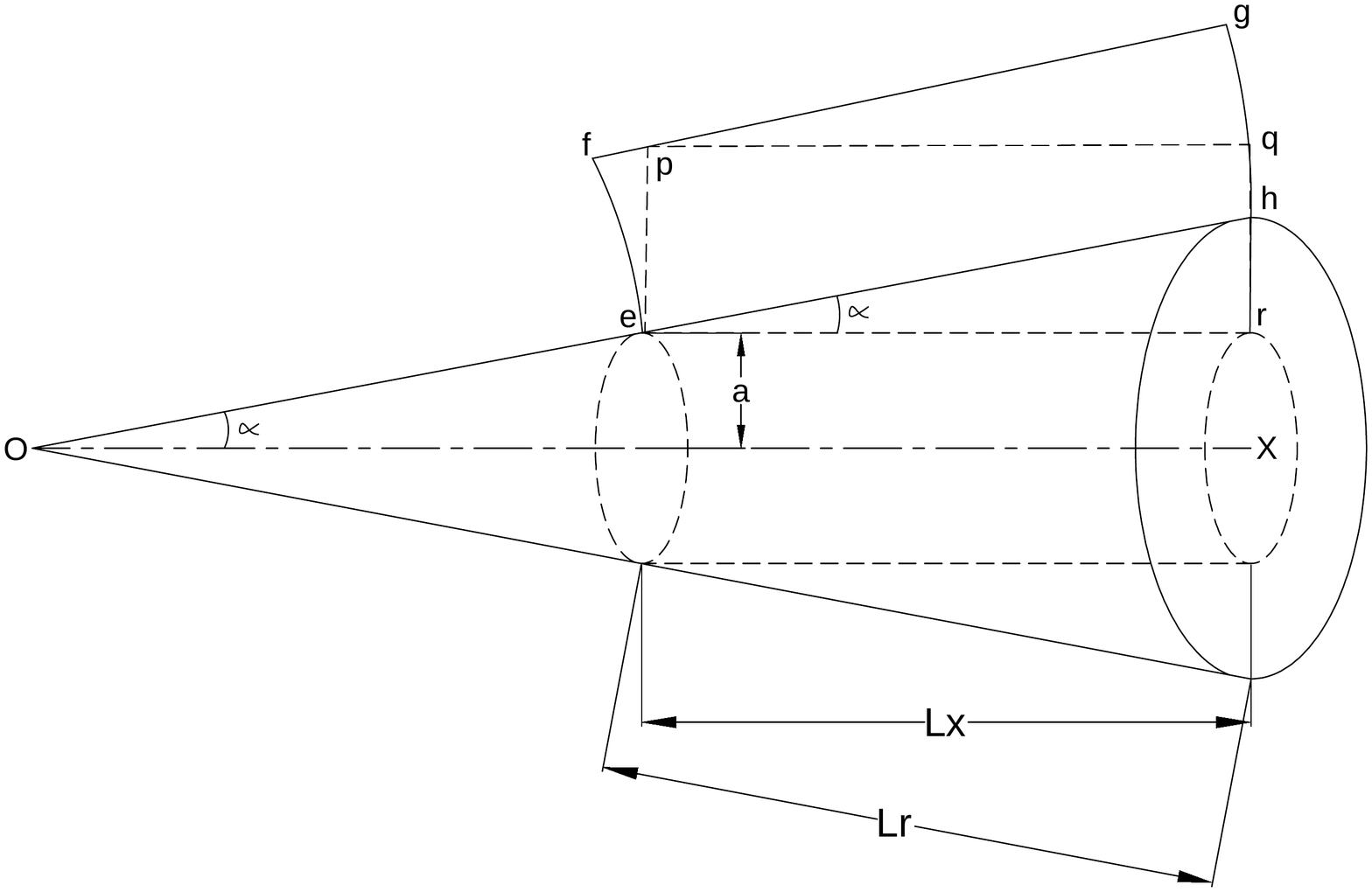}  
\caption{\label{validationcone} Computational domain for Global stability analysis
of axisymmetric boundary layer over a circular cone and cylinder for very small 
semi-cone angle $\alpha$. The domain is modeled in cylindrical (dashed-line) 
and spherical (solid line) coordinates for flow over cylinder and cone respectively.}
\end{center}
\end{figure}

To validate the Global stability results for the incompressible flow over a
circular cone, a blunt cone with the very small semi-cone  angle, 
$\alpha=10^{-12}$ is considered.
Thus the geometry of the circular cylinder and a blunt is nearly similar 
due to very small semi-cone angle $\alpha$ as shown in figure \ref{validationcone}.
A rectangle {\em epqr} shows computational domain for the cylindrical coordinates
and {\em efgh} for spherical coordinates. $L_x$ and $L_r$ are the stream-wise domain length 
for cylindrical  and spherical coordinates.
The stability equations for the axisymmetric flow over circular cylinder and  
cone are written in the polar cylindrical and spherical coordinates respectively.
The Reynolds number is computed based on the body radius (a) at inflow and free-stream 
velocity for both the case. 
The stream-wise and wall normal extent of the domain is also same for both the domain. 
The homogeneous Dirichlet and linear extrapolated boundary conditions are imposed 
at inflow and outflow for both the case.
No-slip and free-stream boundary conditions are applied as usual in wall normal 
direction.
The Global stability results are already validated for the axisymmetric boundary
layer over a circular cylinder \cite{Ramesh}.
\begin{figure}
\begin{center}
\vspace{12pt}
\includegraphics [height=1.75in,width=2.25in, angle=0]
                 {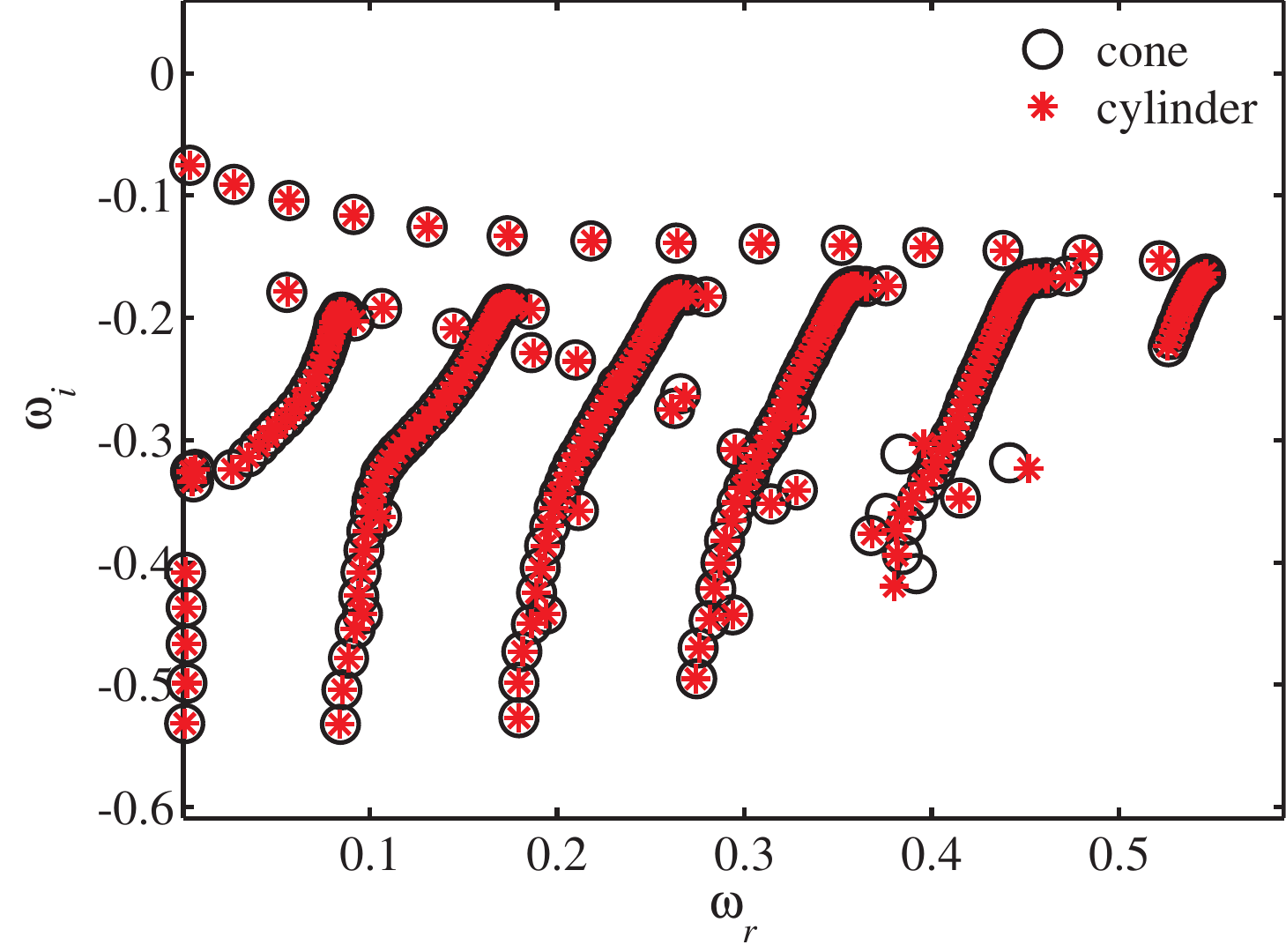}  
\caption{\label{spconecyl} Eigenspectrum comparison of the Global stability 
analysis for the incompressible boundary layer flow over the circular cylinder 
and cone. The Reynolds  number  is Re=1000  based on the body radius.}
\end{center}
\end{figure}
\begin{figure}
\begin{center}
\includegraphics [height=1.75in,width=1.5in, angle=0]
                 {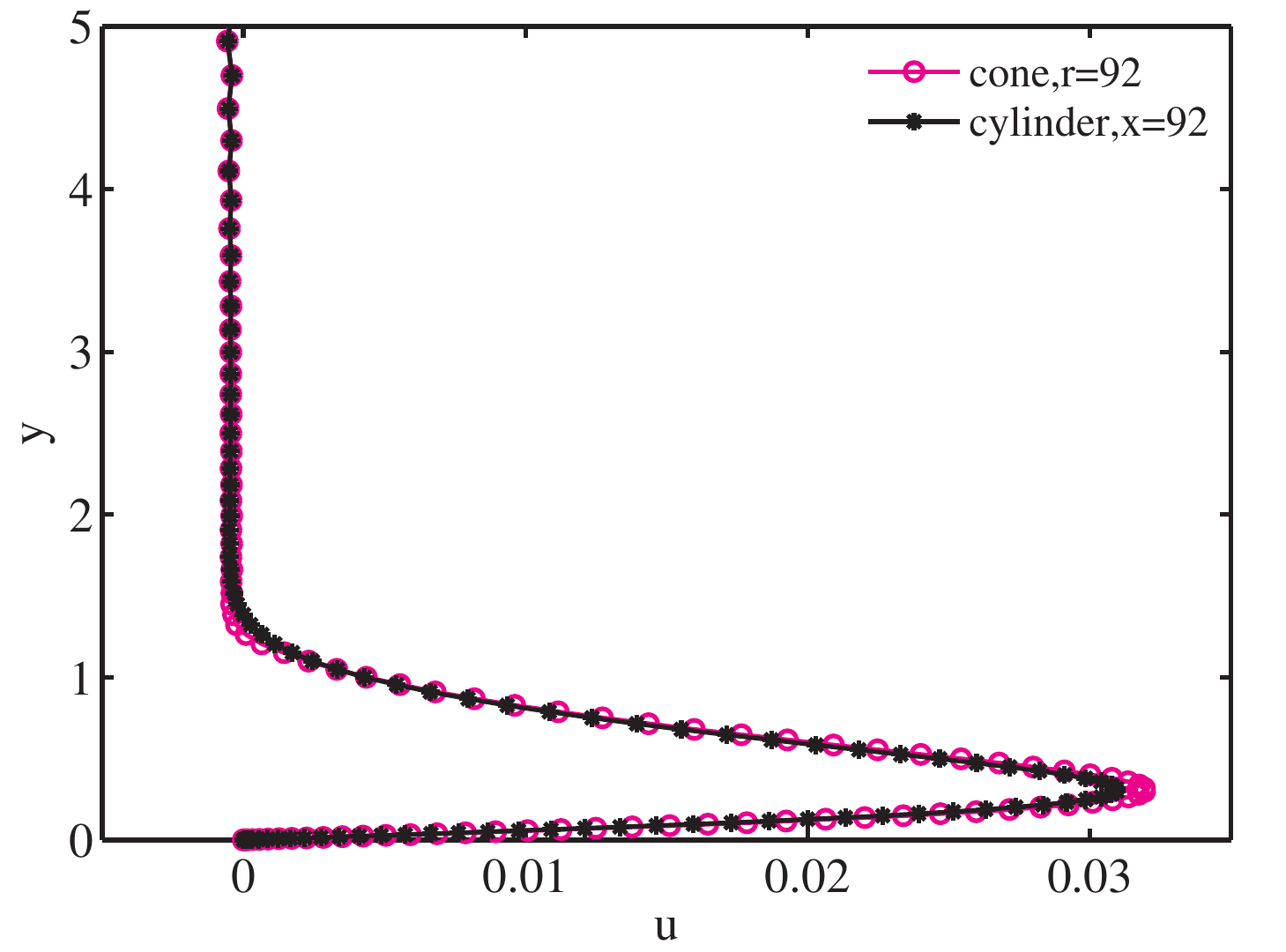}  
\includegraphics [height=1.75in,width=1.5in, angle=0]
                 {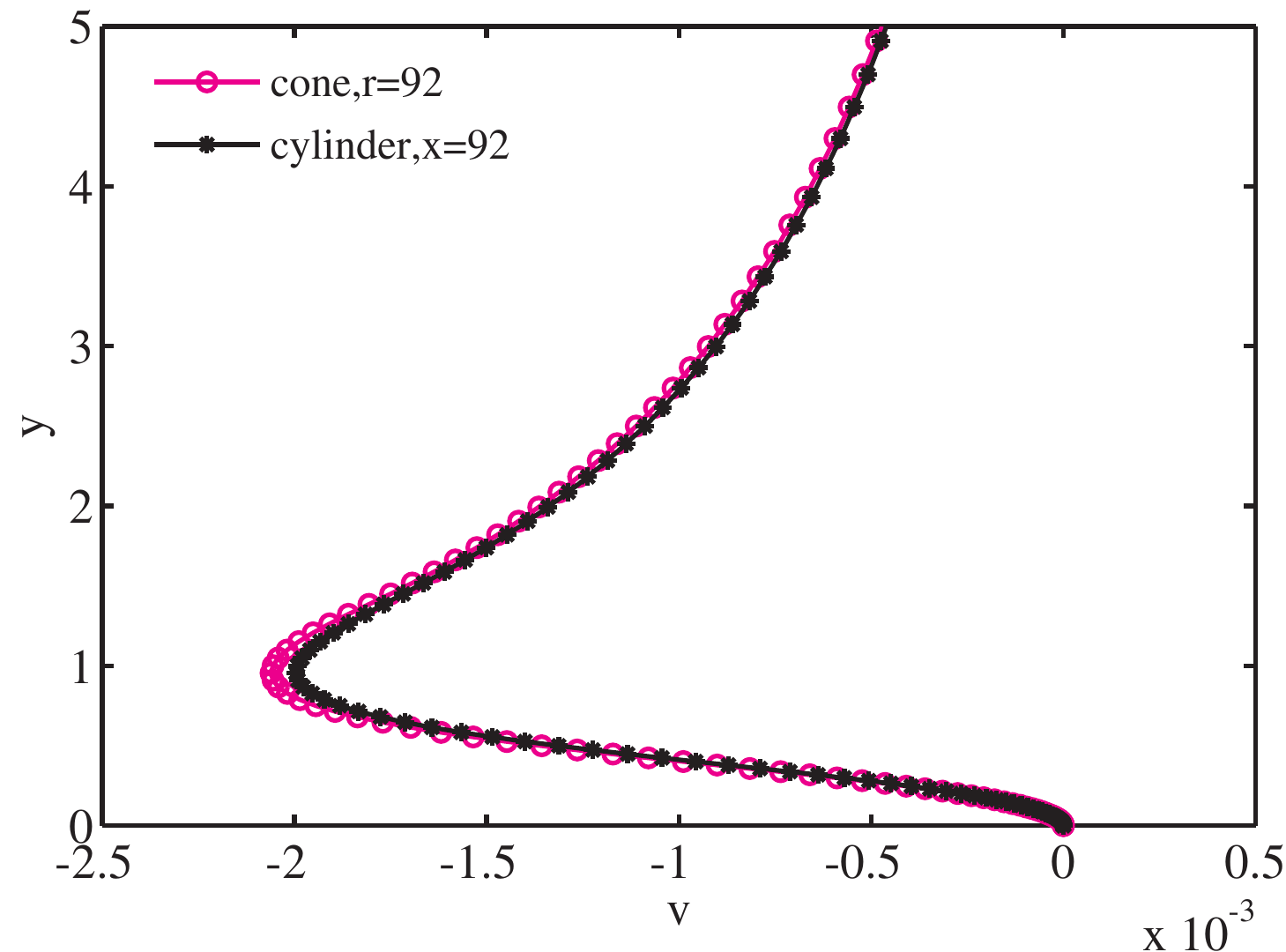} 
\caption{\label{efconecyl} Eigenfunction comparison of the Global stability
analysis for the incompressible boundary layer flow over the circular cone 
and cylinder for Re=1000 at stream-wise distance 92 (a)stream-wise disturbances, 
u and (b) wall normal disturbances, v. Here u and v are stream-wise and wall 
normal disturbances for both cylindrical and spherical coordinates.} 
\end{center}
\end{figure}
Figure \ref{spconecyl} shows the comparison of the eigenspectrum for the
Global stability analysis of the  axisymmetric boundary layer over a circular 
cone and cylinder for $N = 0$ and $Re = 1000$. The discrete and continuous part of 
the spectrum is in excellent agreement with each other. 
Figure \ref{efconecyl} shows the comparison of the eigen-functions for the 
least stable eigenmode at the same stream-wise location. The eigen-functions are 
also in good agreement with each other. 
\section{\label{sec:level1} Results and discussions}
In the present analysis Reynolds number is varying from 174 to 1047 and 
azimuthal wave-numbers from 0 to 5 for different semi-cone angles $\alpha$.
The Reynolds number is computed based on the cone radius(a) at inlet and 
free-stream velocity ($U_\infty$). 
The stream-wise length is normalized by cone radius(a).
The domain size in polar(wall normal) direction is taken as $L_{\theta}=12^o$.
The stream-wise(radial) length is taken 0.75m.
The number of collocation points taken in radial and polar 
directions are n=121 and m=121 respectively. 
The general eigenvalues problem is solved using Arnoldi's iterative algorithm.
The computed eigenvalues are accurate upto three decimal point.
The additional lower resolution cases were also run to verify the 
monotonic convergence of the solution.
Arnoldi's iterative algorithm is used to solve the general eigenvalues problem.
Heavy sponging is applied at the outflow boundary to prevent spurious reflection.
The selected Global eigenmodes are also checked for the spurious mode. 
\begin{table}
 \caption{\label{gcstab} The Grid Convergence study for two leading 
 eigenvalues $\omega_{1}$ and $\omega_{2}$ for $R_{e}=349$,azimuthal 
 wave  number  $N=1$ and semi-cone angle $\alpha=2^o$ for different grid size.
 The grid refinement ratio in each direction is 1.14.
 The maximum relative error is shown here. }   
\begin{ruledtabular}
\begin{tabular}{ccccccc}
 Mesh & n$\times$m & $\omega_{1}$ & $\omega_{2}$ & error(\%) \\ \\
\hline
 \# 1  & 121$\times$121 & 0.01528-0.03609i & 0.02869-0.04375i & 0.174\\
 \# 2  & 107$\times$107 & 0.01528-0.03610i & 0.02874-0.04374i & 0.243\\
 \# 3  & 93$\times$93 & 0.01530-0.03616i & 0.02881-0.04379i & -  \\
\end{tabular}
\end{ruledtabular}
\footnotetext{
The dimensions of computational domain in spherical coordinates 
are $L_r=214.9$ and $L_{\theta}=12^o$ for Global stability computations.
}
\end{table}
A grid  convergence study was performed to check the accuracy level of the
solution and appropriate grid size. 
Table \ref{gcstab} shows the values of two leading eigenvalues computed for $Re=349$, 
$N=1$ and semi-cone angle $\alpha=2^o$ using three different grid size.
The grid resolution was successively improved by a factor of 1.14 in radial and 
polar direction  respectively.
The real and imaginary parts of the eigenvalues shows monotonic convergence of the 
solution with the increased resolution.
In the table \ref{gcstab} n and m indicates  number  of collocation points in 
the radial and polar directions respectively. 
The relative error $\epsilon$ is computed between two consecutive grid sizes for
real and imaginary parts. 
The largest associated error among both the eigenmode is considered.
The mesh 1 is used for all the computations for stability analysis  results 
reported here.
\subsection{\label{sec:level1} Semi-cone angle $\alpha=2^o$}
\begin{figure}
\begin{center}
\vspace{12pt}
\includegraphics[height=1.75in,width=2.25in, angle=0]
                {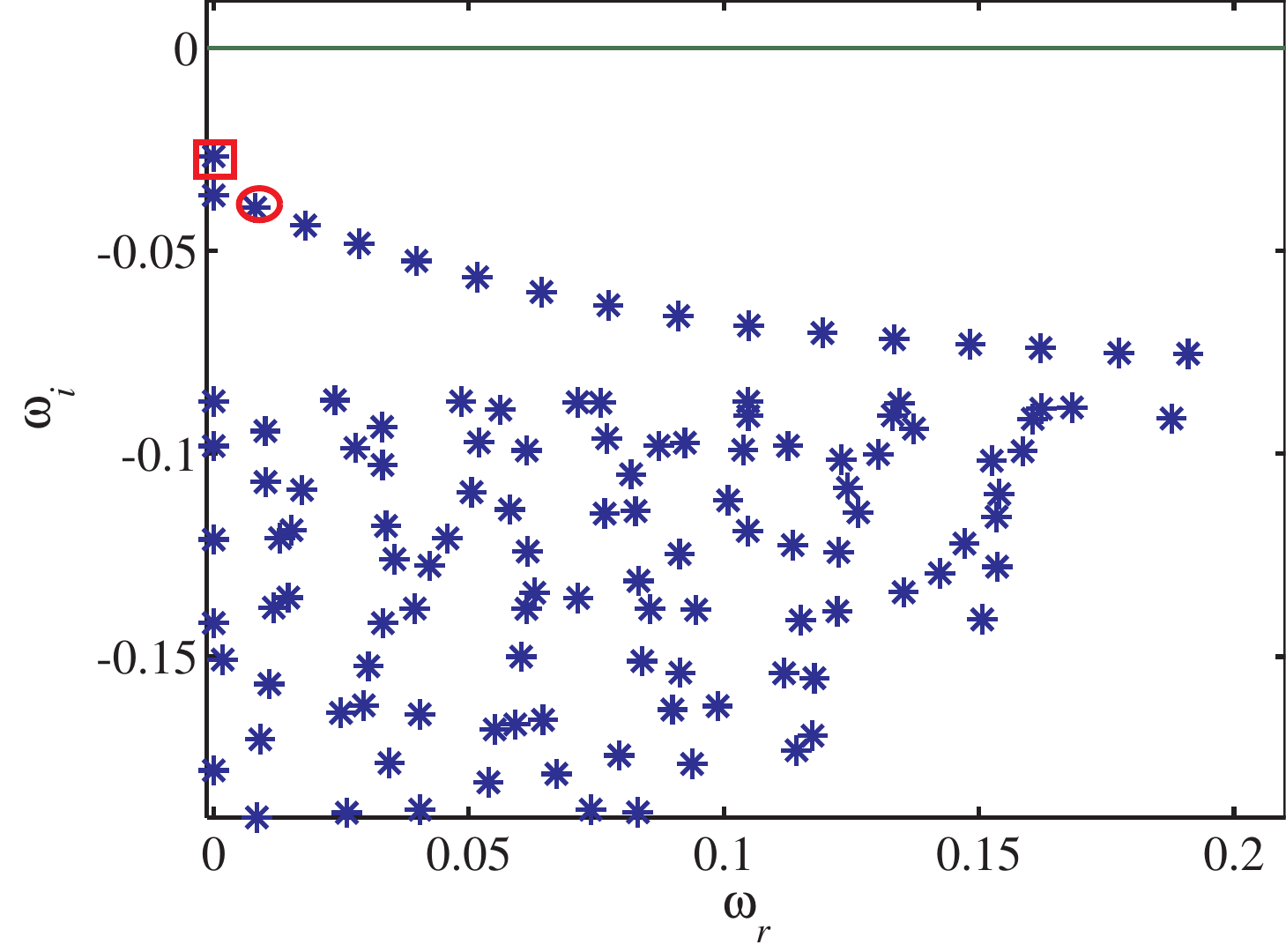}  
\caption{\label{sp2deg} Eigenspectrum for axisymmetric mode (N=0) 
         and Re=349 for semi-cone angle $\alpha=2^o$.}
\end{center}
\end{figure}
\begin{figure}
\begin{center}
\includegraphics [height=1.25in,width=1.65in, angle=0]
                 {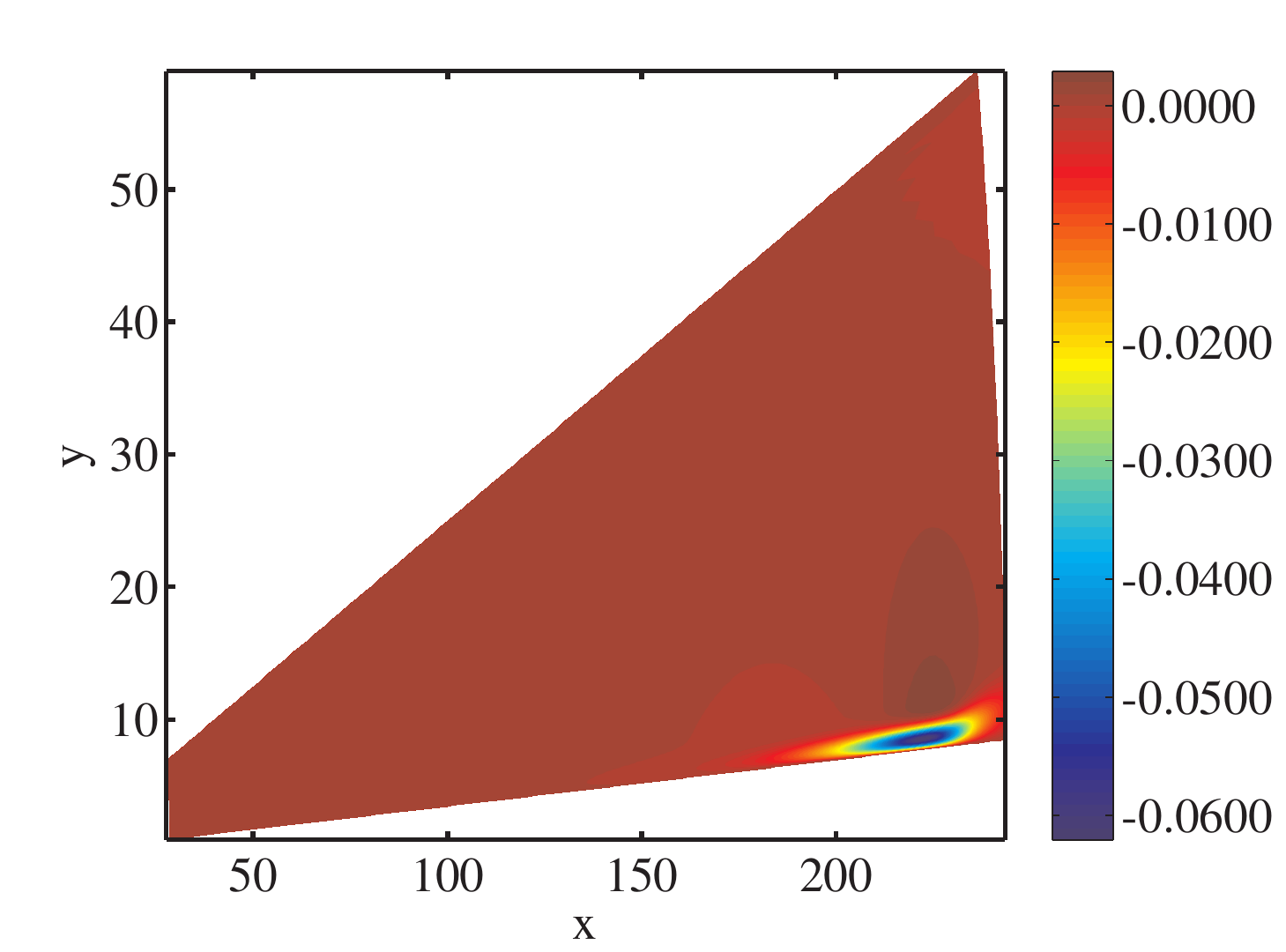}  
\includegraphics [height=1.25in,width=1.65in, angle=0]
                 {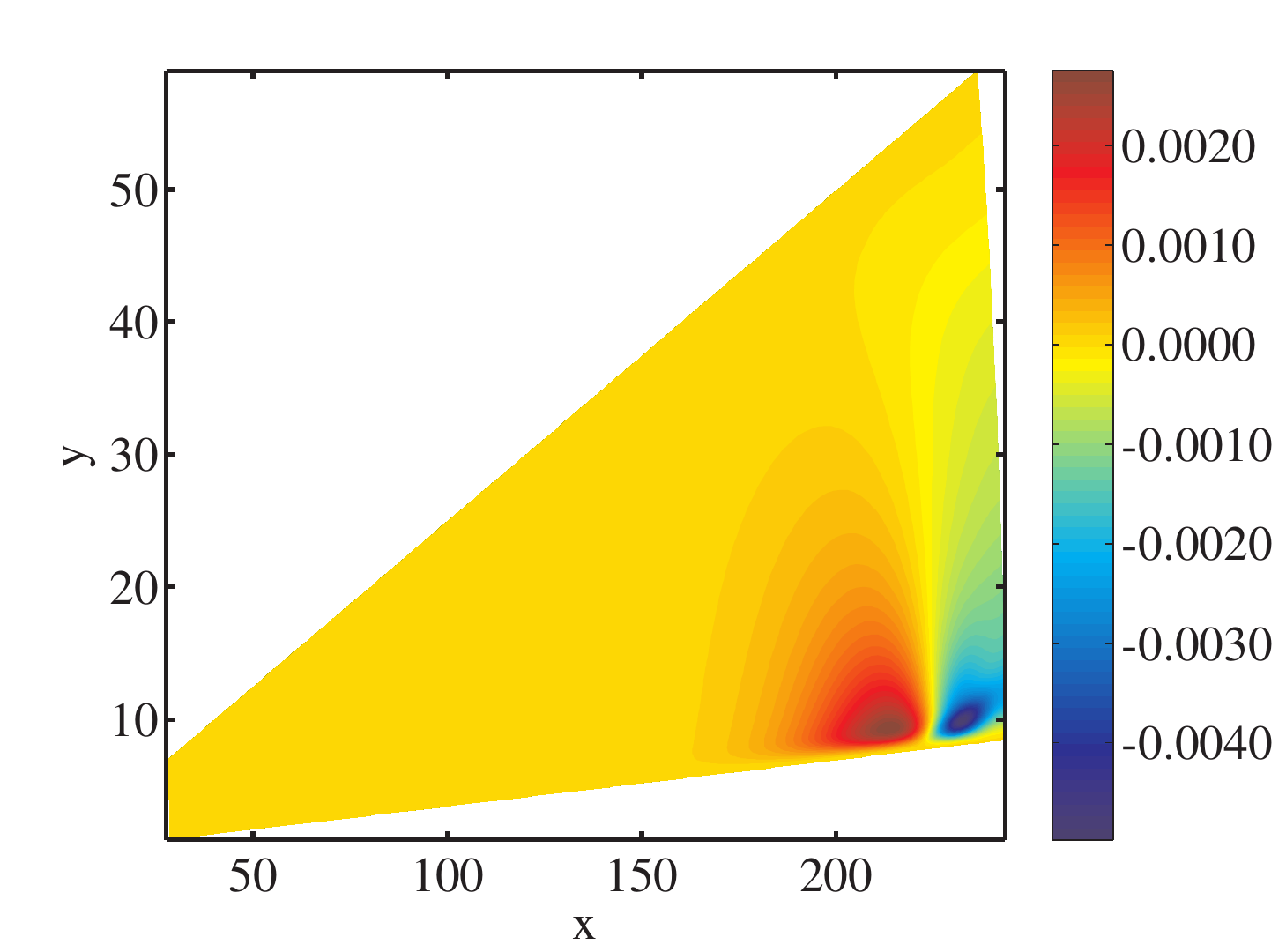} 
\caption{\label{ef2degstn} Contour plot of the real parts of 
(a) stream-wise $u_{r}$ and (b) wall normal  $u_{\theta}$ velocity disturbances
for stationary eigenmode, $\omega=0-0.02667i$ for semi-cone angle $\alpha=2^o$ 
marked by square in the figure \ref{sp2deg}.} 
\end{center}
\end{figure}
\begin{figure}
\begin{center}
\includegraphics [height=1.25in,width=1.65in, angle=0]
                 {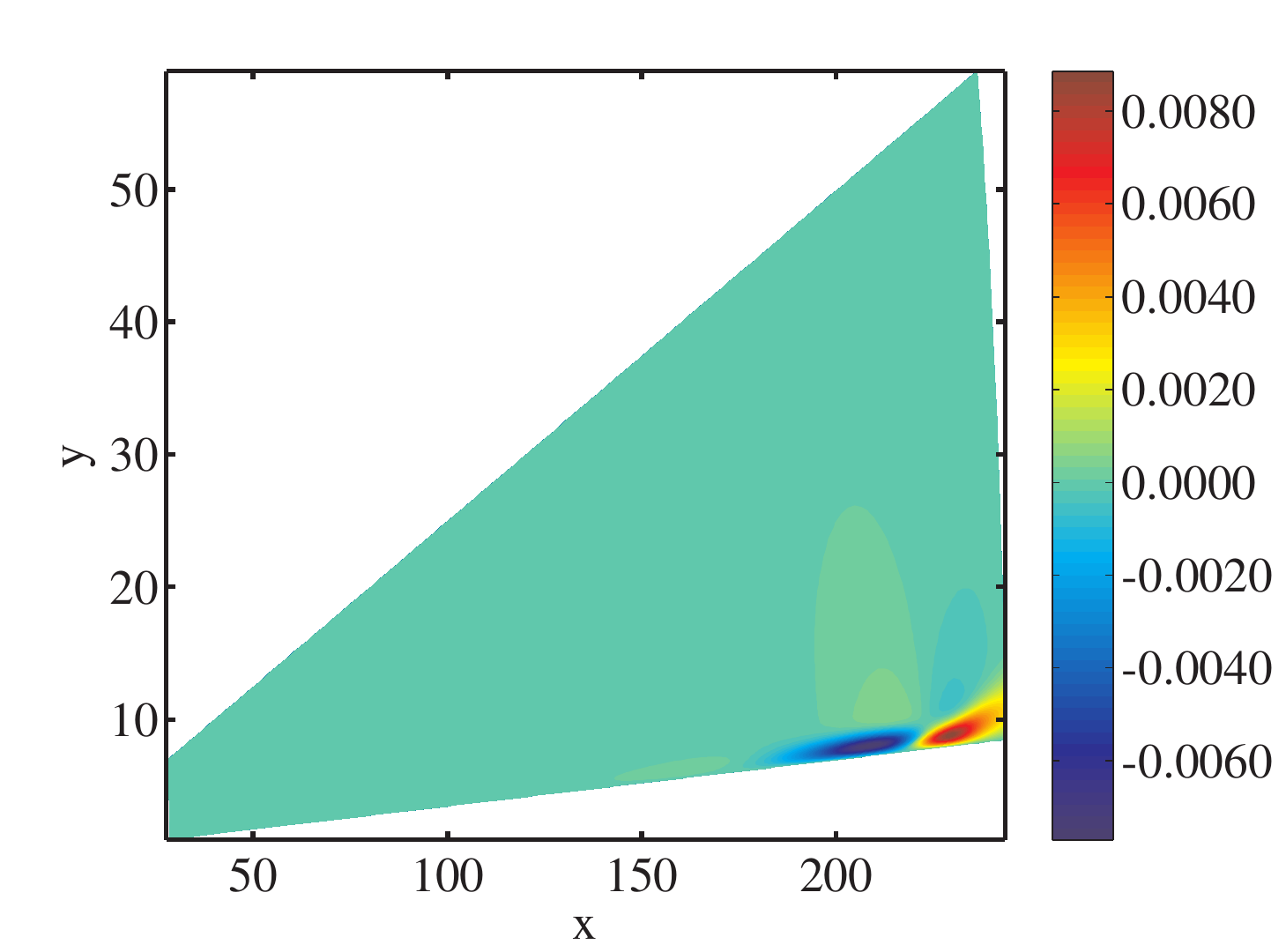}  
\includegraphics [height=1.25in,width=1.65in, angle=0]
                 {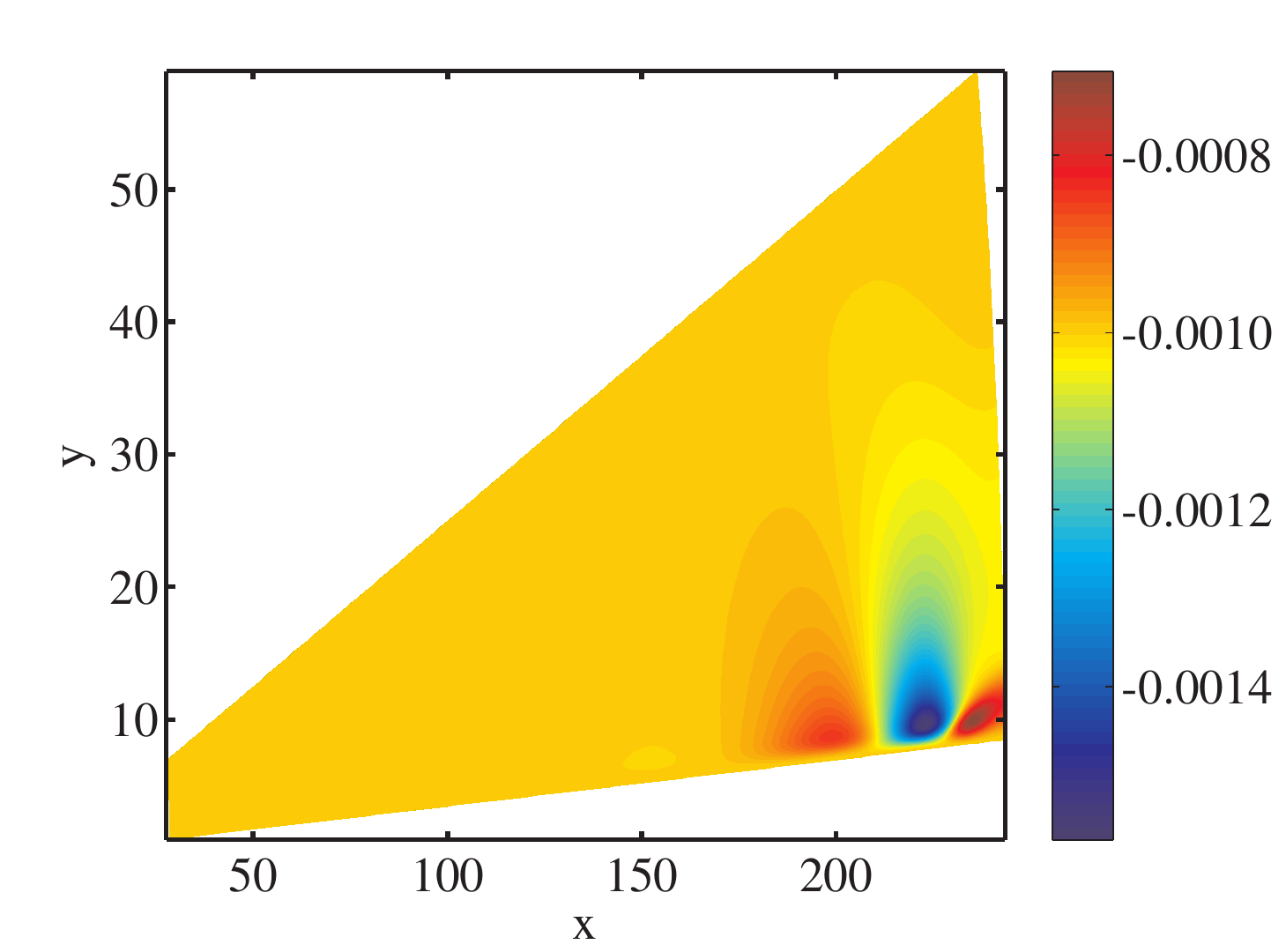} 
\caption{\label{ef2degosc1} Contour plot of the real parts of 
(a) stream-wise $u_{r}$ and (b) wall normal  $u_{\theta}$ velocity disturbances
for oscillatory eigenmode, $\omega=0.008154-0.03937i$ for semi-cone angle $\alpha=2^o$ 
marked by ellipse in the figure \ref{sp2deg}.} 
\end{center}
\end{figure}
\begin{figure}
\begin{center}
\includegraphics [height=1.25in,width=1.65in, angle=0]
                 {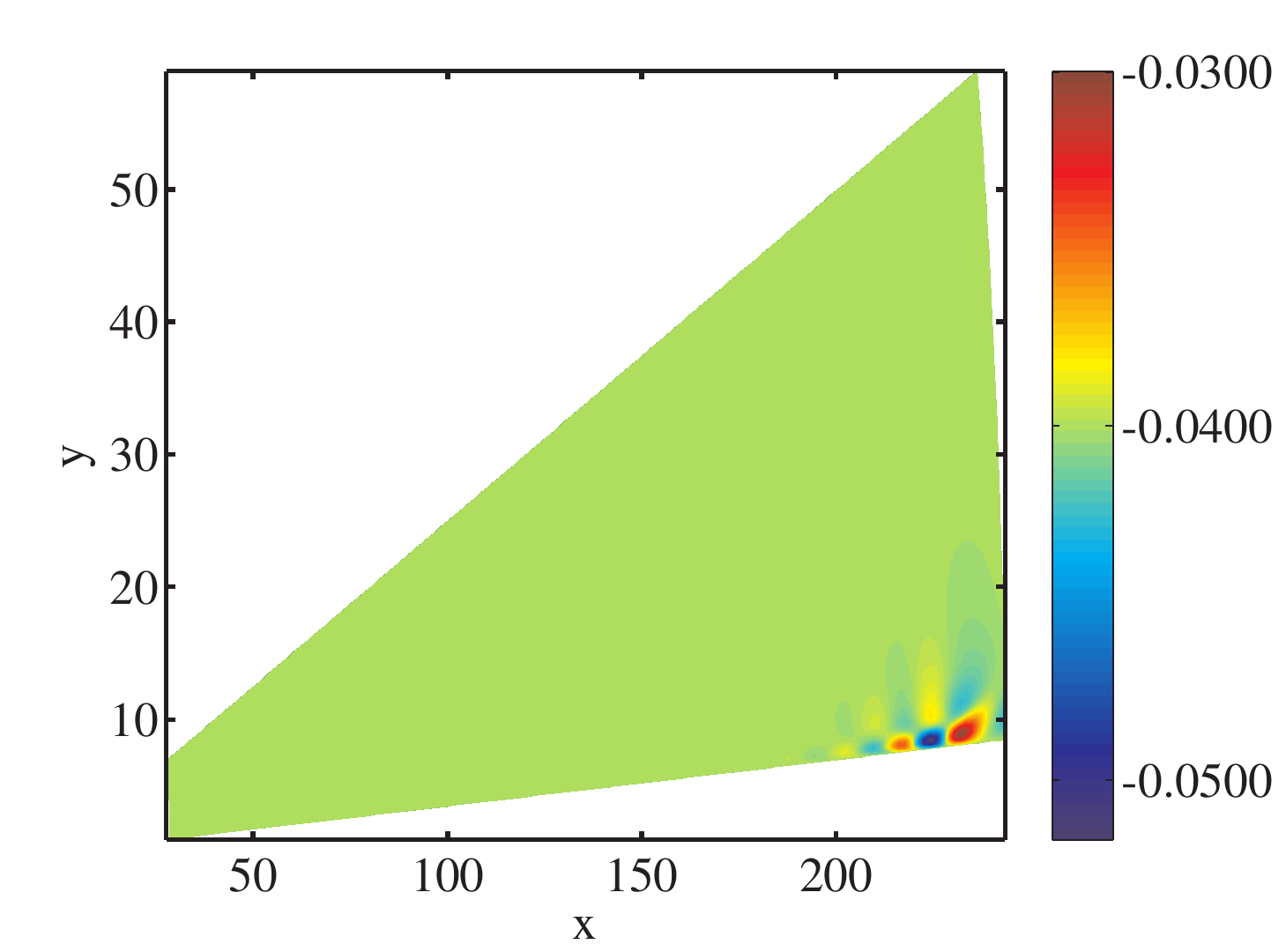}  
\includegraphics [height=1.25in,width=1.65in, angle=0]
                 {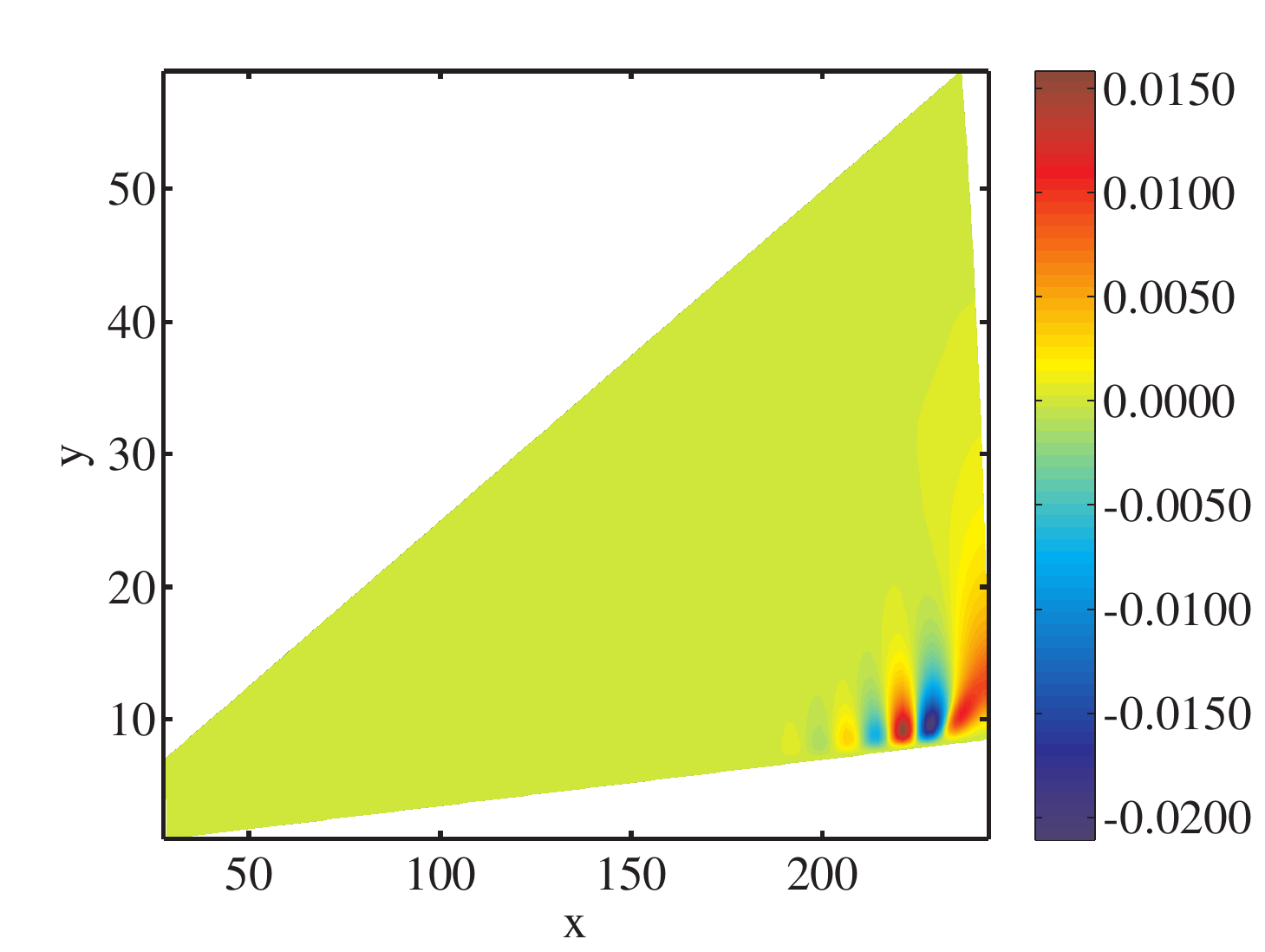} 
\caption{\label{ef2degosc2} Contour plot of the real parts of 
(a) stream-wise $u_{r}$ and (b) wall normal $u_{\theta}$ velocity disturbances
for oscillatory eigenmode, $\omega=0.1775-0.07512i$ for semi-cone angle $\alpha=2^o$.} 
\end{center}
\end{figure}

Figure \ref{sp2deg} shows the eigenspectrum of axisymmetric mode ($N=0$)
for $Re=349$ and semi-cone angle $\alpha=2^o$.
The eigenmodes marked by square and circle are called stationary and 
oscillatory mode.
The stationary mode has a complex frequency $\omega=0-0.02667i$.
The Global mode is temporally stable because $\omega_{i} < 0$ and hence the 
amplitudes of the disturbances decay in time.
Figure \ref{ef2degstn}(a) and (b) presents the two dimensional spatial 
structure of the eigenmodes for radial ($u_{r}$) and polar ($u_{\theta}$)
velocity disturbances. 
The magnitudes of the disturbance amplitudes are zero at inlet as it is applied 
inlet boundary condition.
The disturbance  amplitudes evolve with the time in the flow domain and moves
in the stream-wise direction towards downstream.
The size and magnitudes of the disturbances grows as they move towards downstream.
The magnitudes of the $u_{r}$ disturbances are one order higher then that of $u_{\theta}$.
The polar disturbances $u_\theta$ contaminates the flow field upto large extent than
that of $u_r$ disturbances, however its magnitudes very small compare to $u_\theta$.
Figure \ref{ef2degosc1} shows the spatial structure of oscillatory eigenmode.
The associated complex frequency for this mode is $\omega=0.008154-0.03937i$.
This oscillatory Global mode is also temporally stable, because $\omega_{i} < 0$. 
The variation of the amplitude is not monotonic for this eigenmode.
The wave-like nature of the disturbances are found for this mode.
The oscillatory modes evolved in the flow field, grows in size and magnitude
while moving towards the downstream and hence the flow is convectively unstable. 
Figure \ref{ef2degosc2} shows the two dimensional mode structure of
the $u_{r}$ and $u_\theta$ for the $\omega_r=0.1775$.
The  stream-wise domain  length is 0.0000 .
The disturbances are observed to evolve near the wall surface and decay exponentially
at free-stream.
The typical length scale of the wavelet structure decreases with the increase 
in the frequency $\omega_r$.
It also demonstrates that the region of contamination reduces with the  increases in frequency. 
The length scale of the wavelet structure is found small for the increased frequency 
as shown in figure \ref {ef2degosc2}.
\subsection{\label{sec:level1} Semi-cone angle $\alpha=4^o$}
\begin{figure}
\begin{center}
\vspace{12pt}
\includegraphics [height=1.75in,width=2.25in, angle=0]
                 {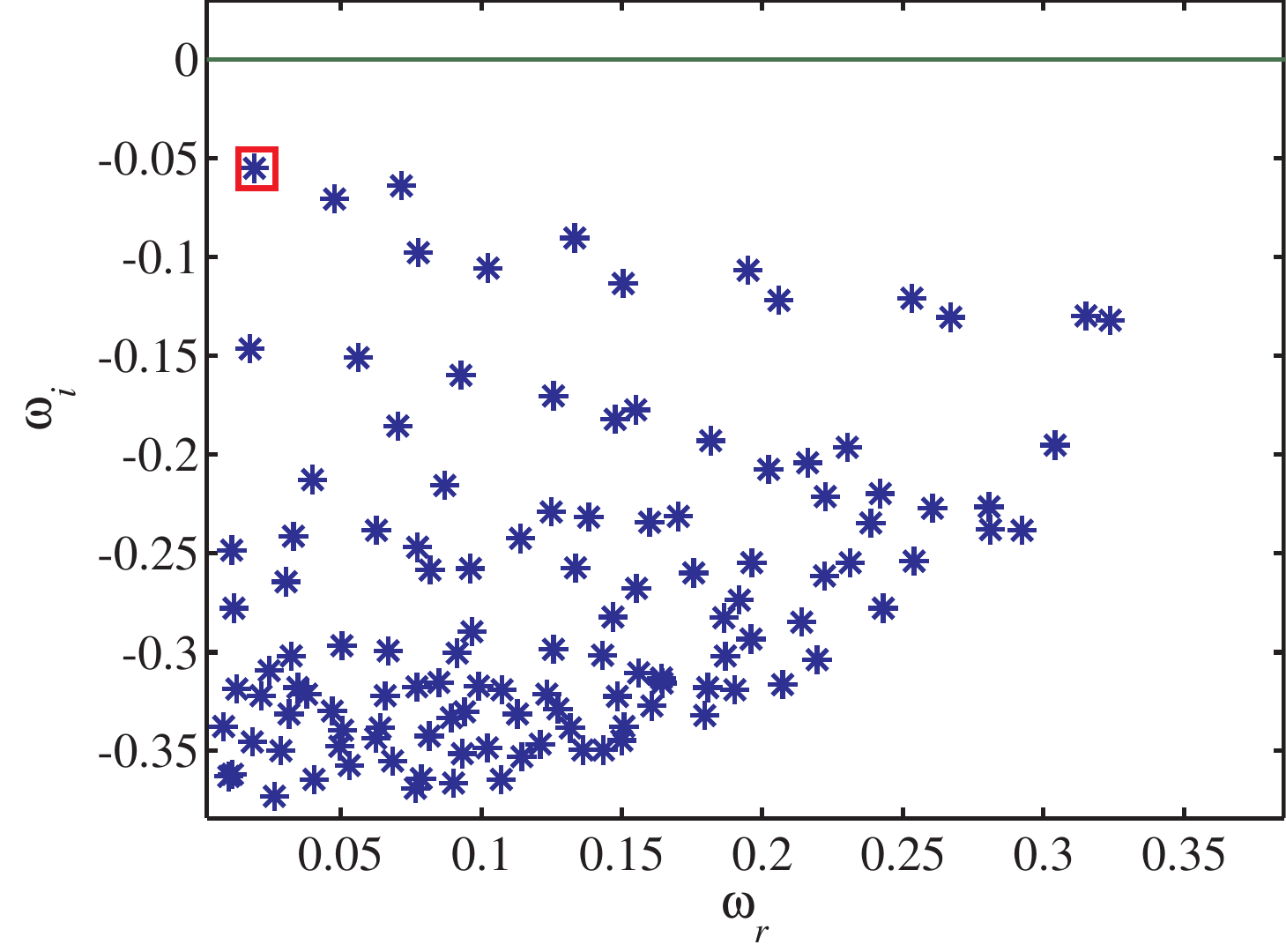}  
\caption{\label{sp4deg} Eigenspectrum for helical mode (N=1) and Re=523 
for semi-cone angle $\alpha=4^o$.}
\end{center}
\end{figure}
\begin{figure}
\begin{center}
\includegraphics [height=1.25in,width=1.65in, angle=0]
                 {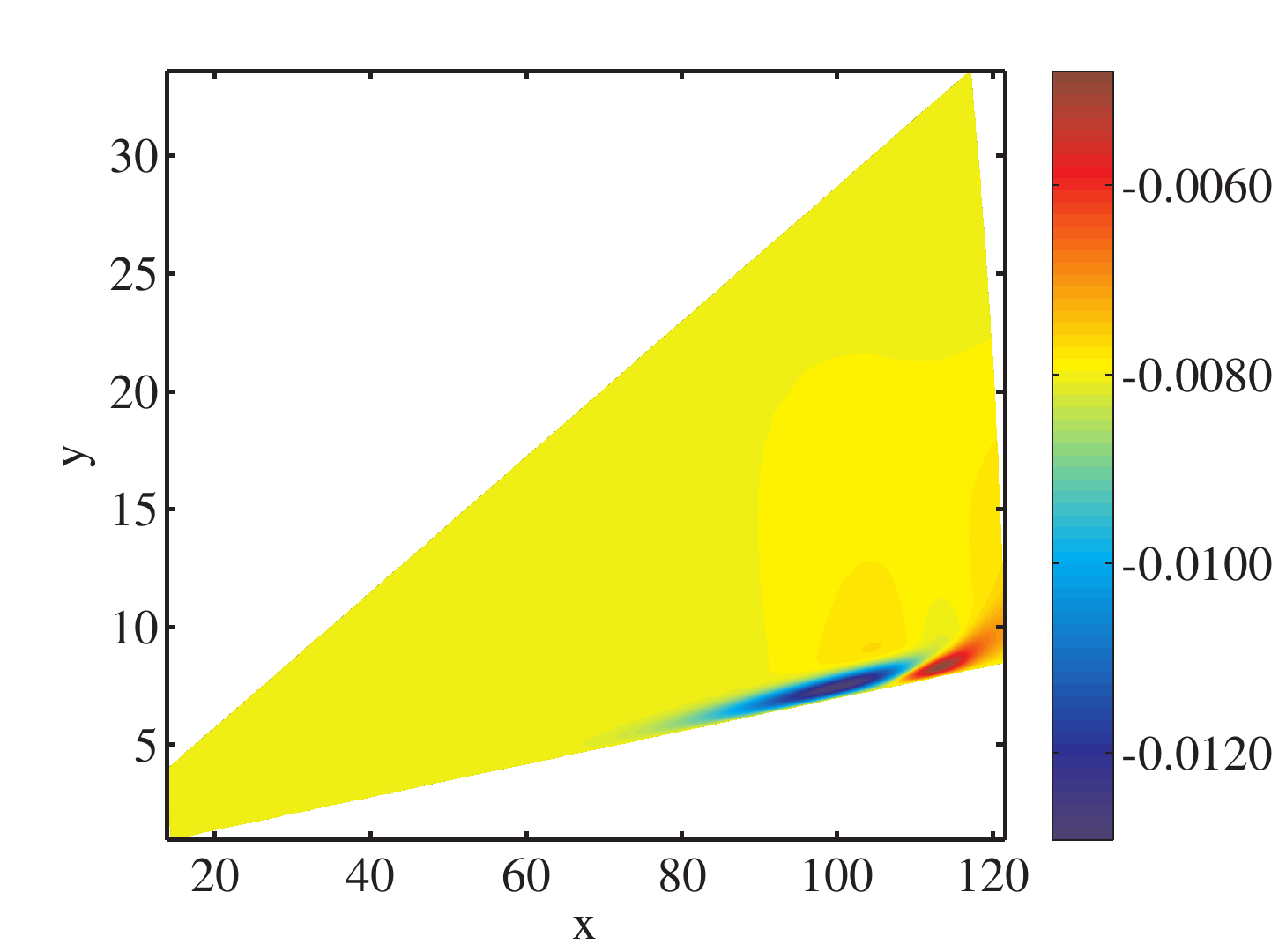}  
\includegraphics [height=1.25in,width=1.65in, angle=0]
                 {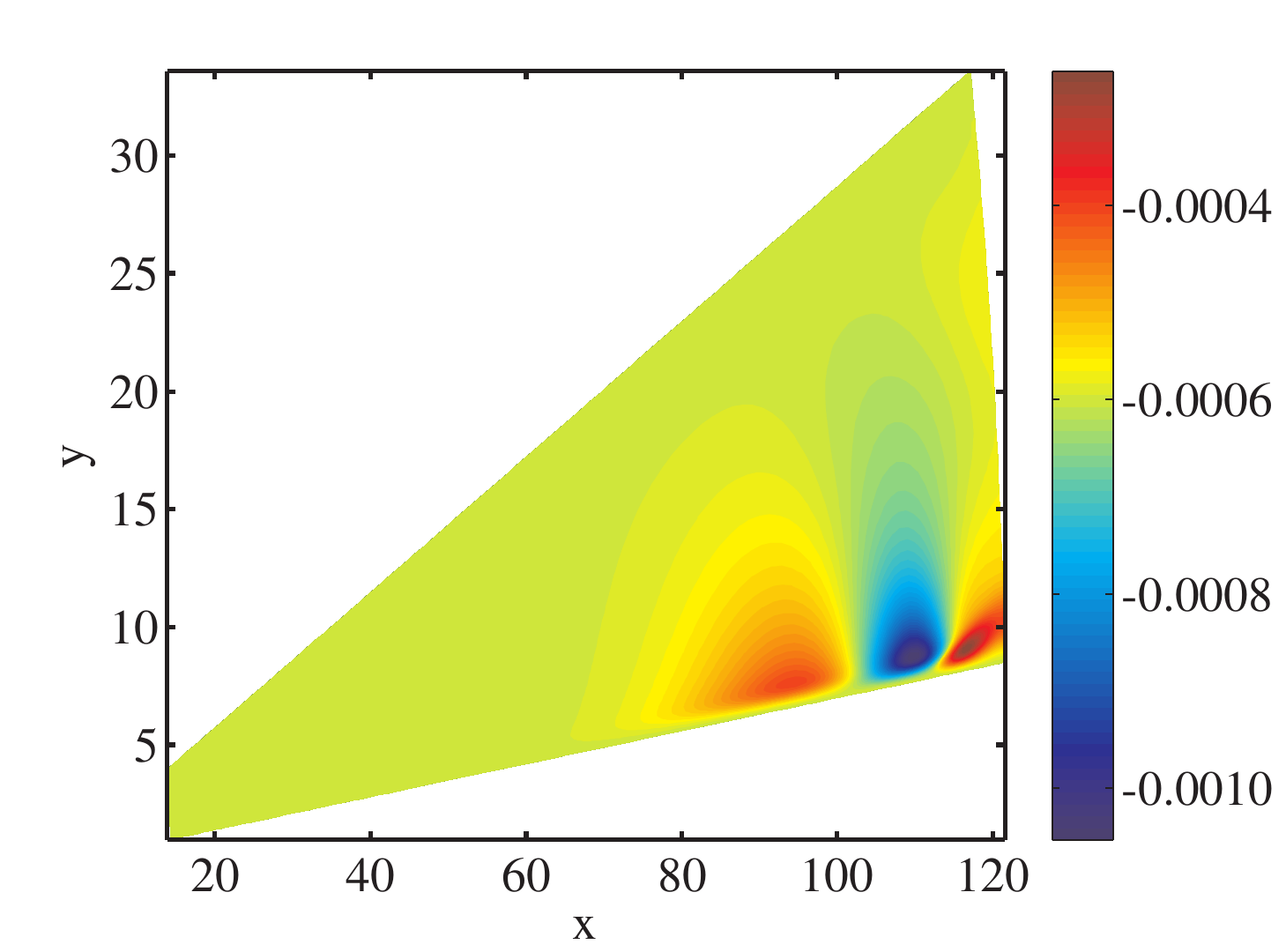} 
\end{center}
\begin{center}
\includegraphics [height=1.25in,width=1.65in, angle=0]
                 {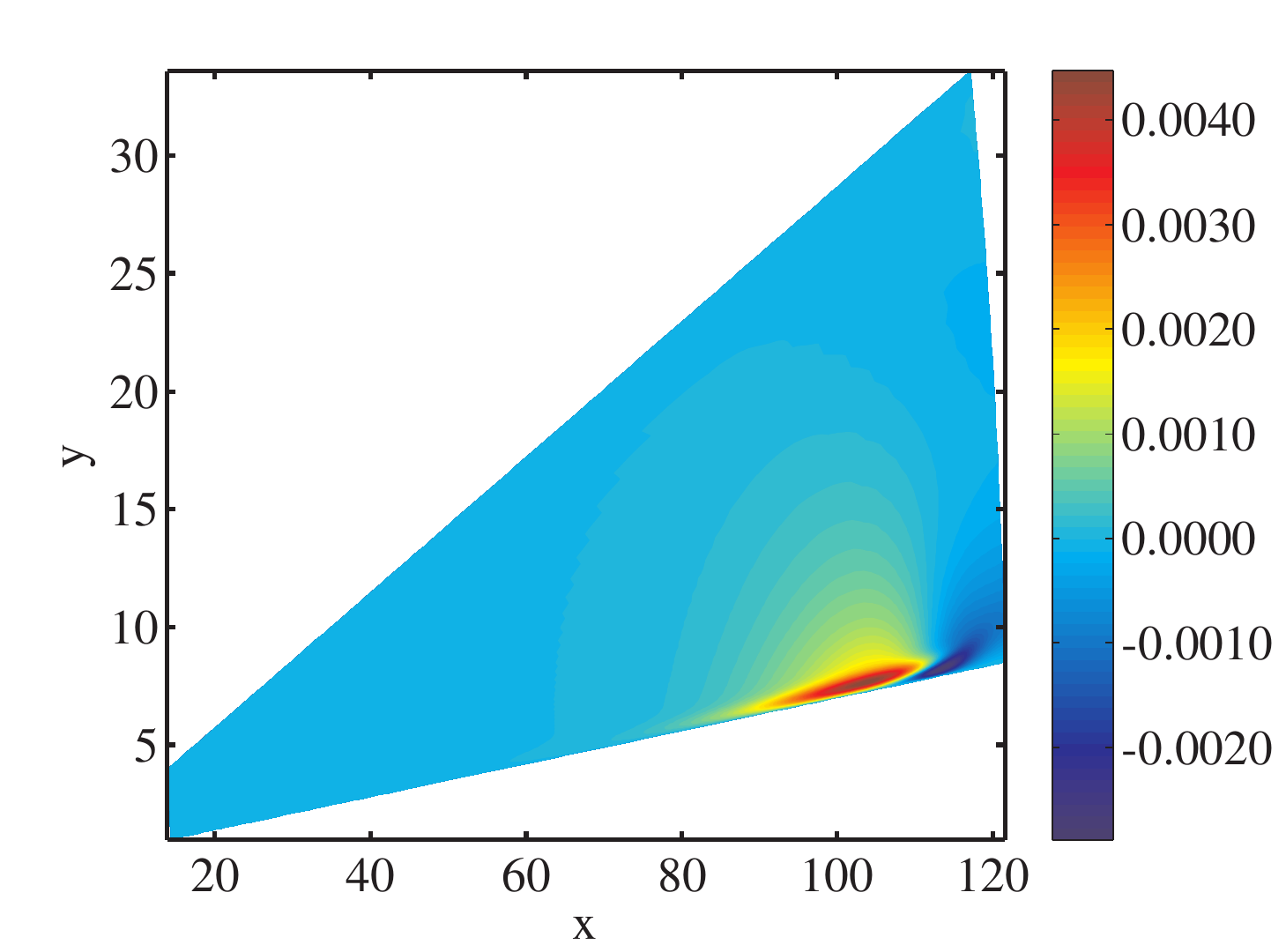}  
\caption{\label{ef4deg} Contour plot of the real parts of 
(a) stream-wise $u_{r}$ and (b) wall normal  $u_{\theta}$ velocity disturbances
for stationary eigenmode, $\omega=0.01942-0.05512i$ for semi-cone angle $\alpha=4^o$ 
marked by square in the figure \ref{sp4deg}.} 
\end{center}
\end{figure}

Figure  \ref{sp4deg} shows the spectrum for helical mode N=1, Re=523 and semi-cone
angle  $\alpha=4^o$.
The most unstable oscillatory mode has an eigenvalue $\omega=0.01942-0.05512i$.
The Global mode is temporally stable because largest $\omega_i<0$.
Figure \ref{ef4deg} shows the two dimensional mode structure for the 
N=1, Re=523 and  semi-cone angle $\alpha=4^o$.
It has been observed that the disturbances evolved in the flow field at the earlier stage 
than that of a cone with semi-cone angle $\alpha=2^o$.  
The magnitudes of the disturbance velocity components and the region of contamination in 
polar direction are higher then that of $\alpha=2^o$, which makes the flow convectively 
more unstable.
The largest $\omega_i$ has reduced with the increased semi-cone angle $\alpha$ for 
a given Reynolds number, which makes the Global modes more stable.
\subsection{\label{sec:level1} Semi-cone angle $\alpha=6^o$}
\begin{figure}
\begin{center}
\vspace{12pt}
\includegraphics [height=1.75in,width=2.25in, angle=0]
                 {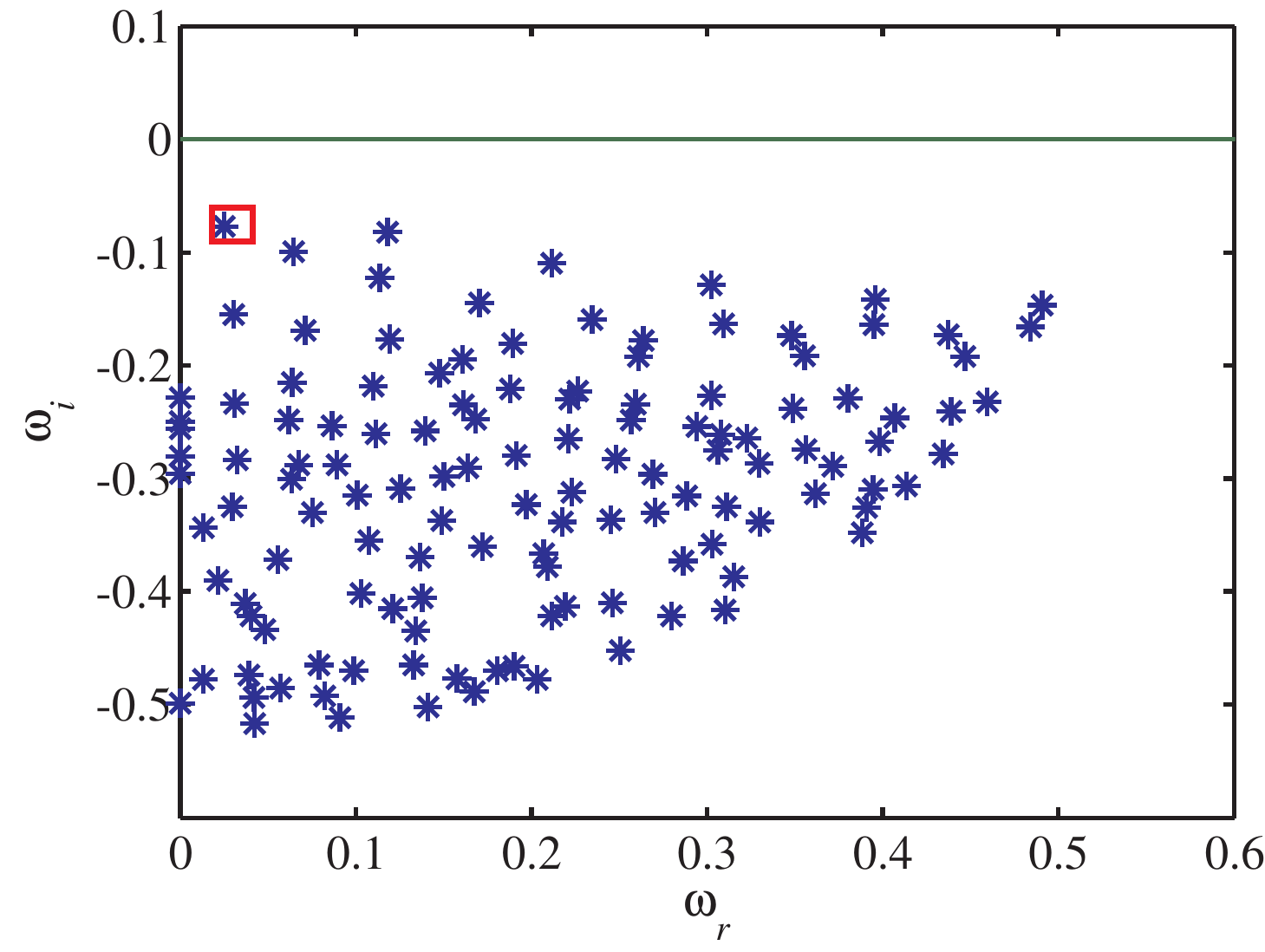}  
\caption{\label {sp6deg} Eigenspectrum for helical mode (N=1) and Re=698 
for semi-cone angle $\alpha=6^o$.}
\end{center}
\end{figure}
\begin{figure}
\begin{center}
\includegraphics [height=1.25in,width=1.65in, angle=0]
                 {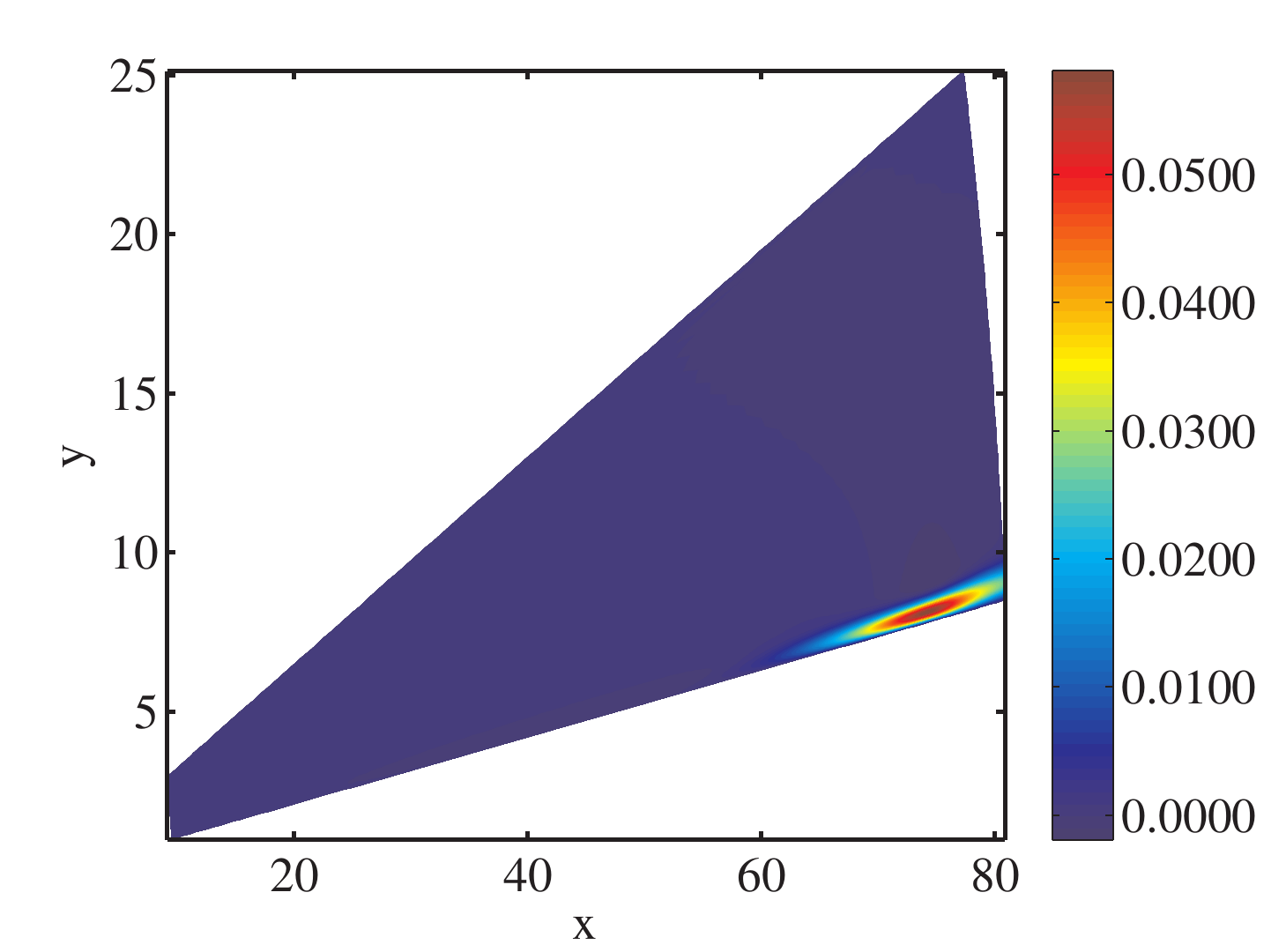}  
\includegraphics [height=1.25in,width=1.65in, angle=0]
                 {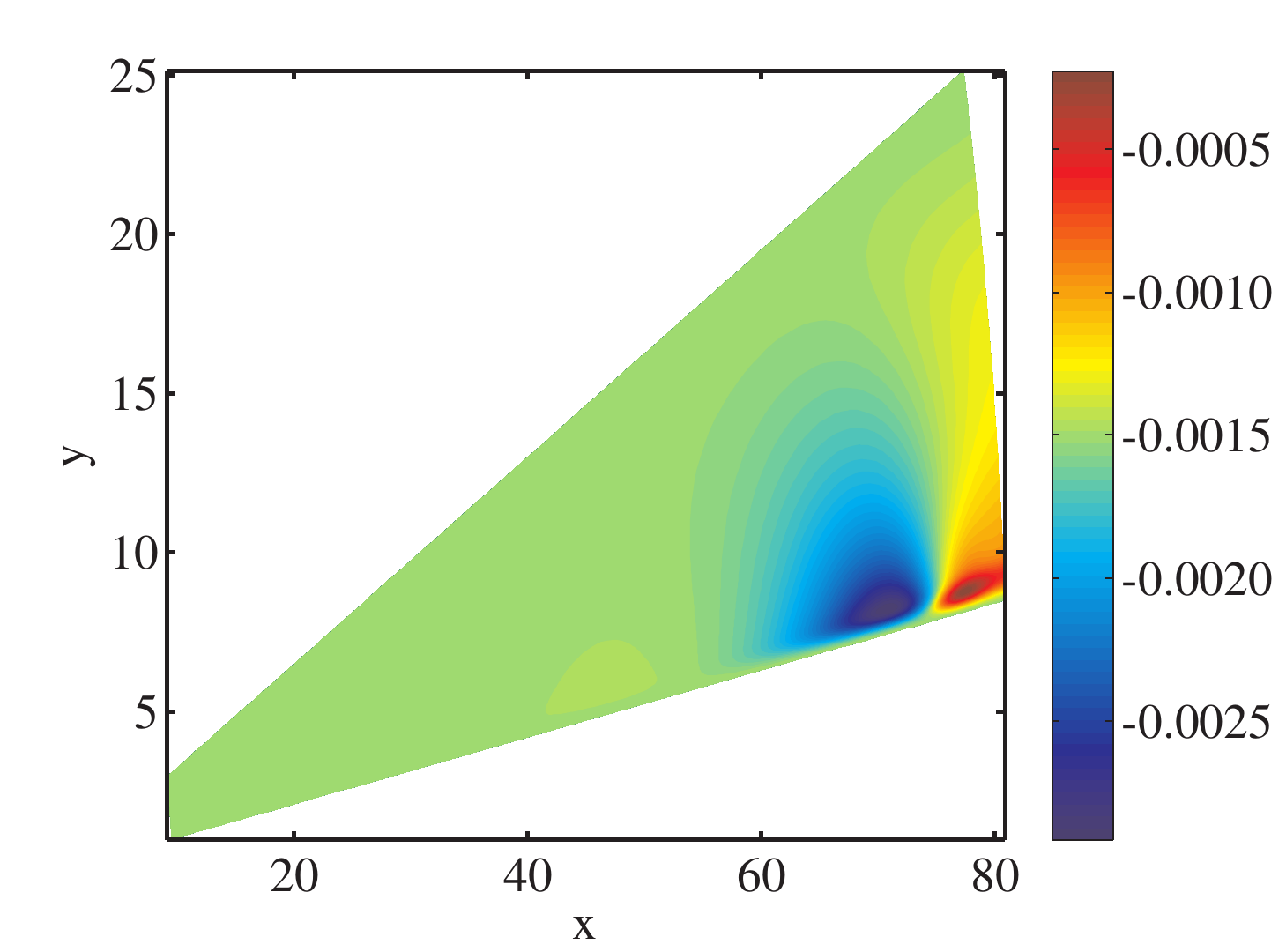} 
\end{center}
\begin{center}
\includegraphics [height=1.25in,width=1.65in, angle=0]
                 {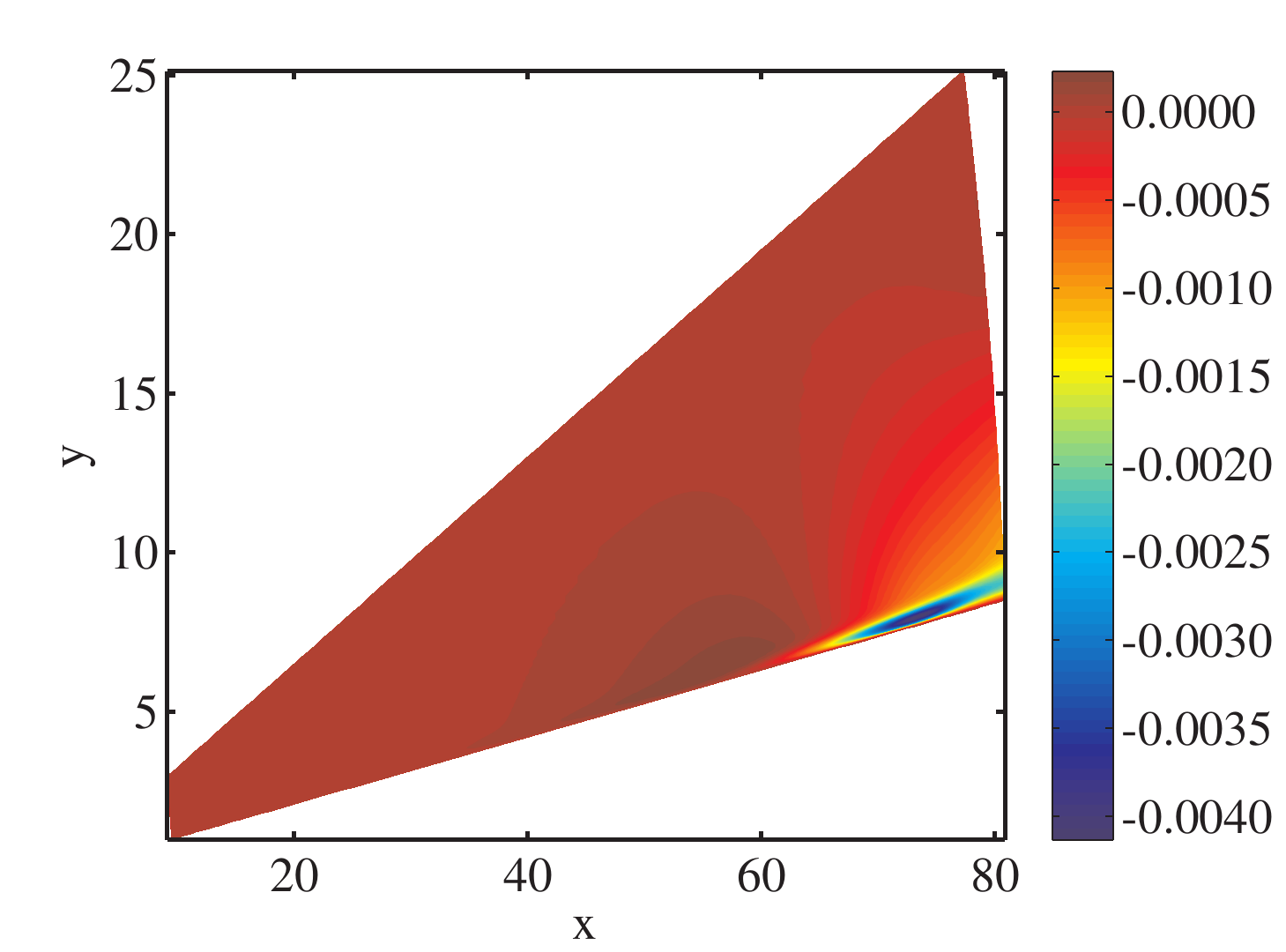}  
\caption{\label{ef6deg} Contour plot of the real parts of 
(a) stream-wise $u_{r}$ and (b) wall normal  $u_{\theta}$ velocity disturbances
for stationary eigenmode, $\omega=0.02539-0.07707i$ for semi-cone angle 
$\alpha=6^o$ marked by square in the figure \ref{sp6deg}.} 
\end{center}
\end{figure}

Figure \ref{sp6deg} shows the spectrum for helical mode N=1, 
Re=698 and semi-cone angle $\alpha=6^o$.
The most unstable oscillatory mode has an eigenvalue $\omega=0.02539-0.07707i$.
The Global mode is temporally stable because largest $\omega_i<0$.
Figure \ref{ef6deg} shows the two dimensional mode structure for the 
N=1, Re=698 and  semi-cone angle $\alpha=6^o$.
The structure of discrete part of the spectrum disturbs with the increase in 
semi-cone angle $\alpha$. 
The magnitudes of the $u_{r}$ disturbance amplitudes is one order higher than that
of $u_{\theta}$ and $u_{\Phi}$.
With the increase in semi-cone angle $\alpha$ the disturbances starts to 
grow at  early  stage.
It has been observed that the disturbances evolved in the flow field at the earlier stage 
than that of a cone with semi-cone angle $\alpha=2^o$.  
The spatial structure of the disturbances are of similar nature, however  damping rate
of the Global mode increases with the increase in semi-cone angle.
\subsection{\label{sec:level1} Temporal growth rate}
\begin{figure}
\begin{center}
\includegraphics[height=1.25in,width=1.65in, angle=0]
                {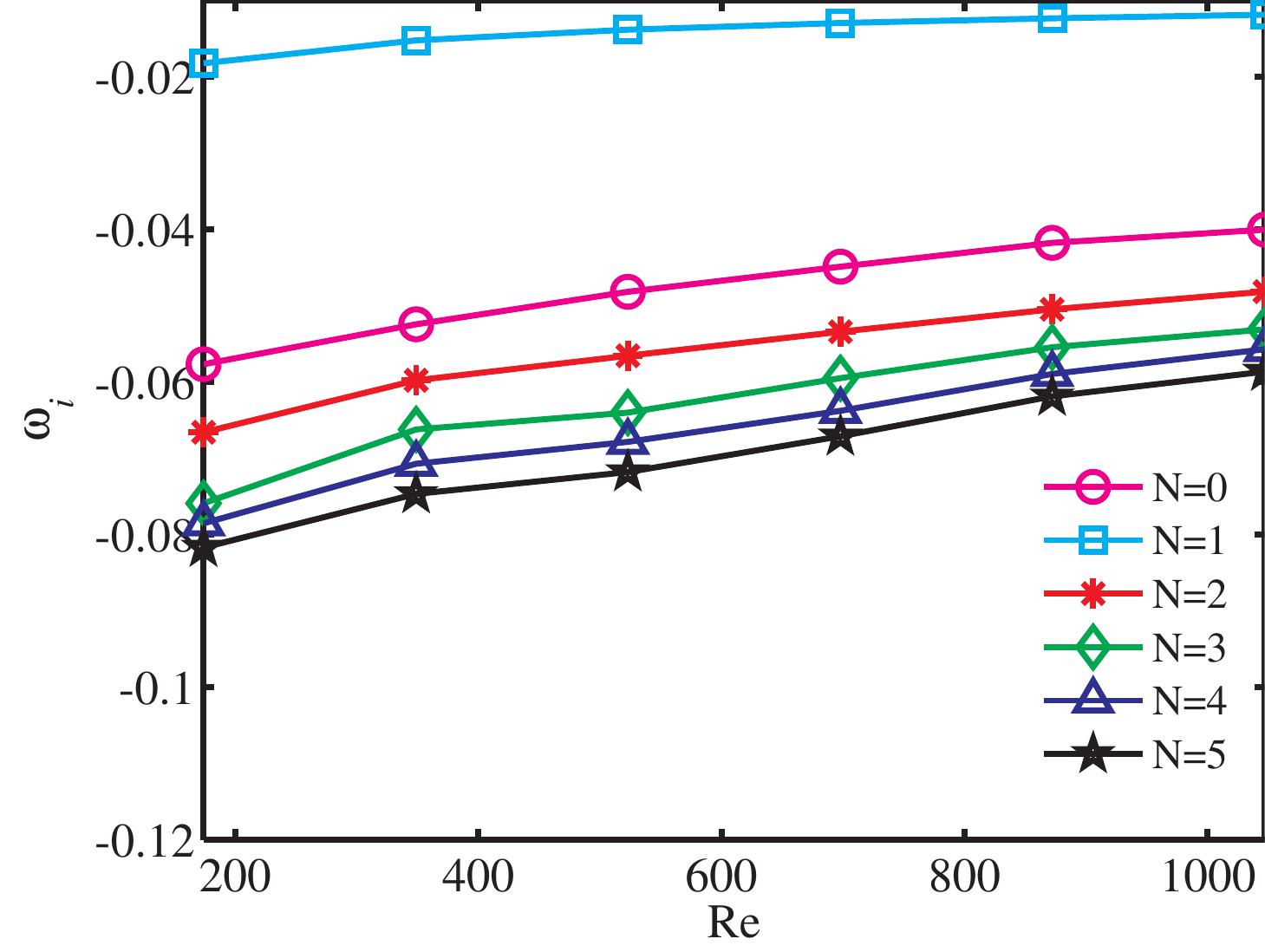}  
\includegraphics[height=1.25in,width=1.65in, angle=0]
                {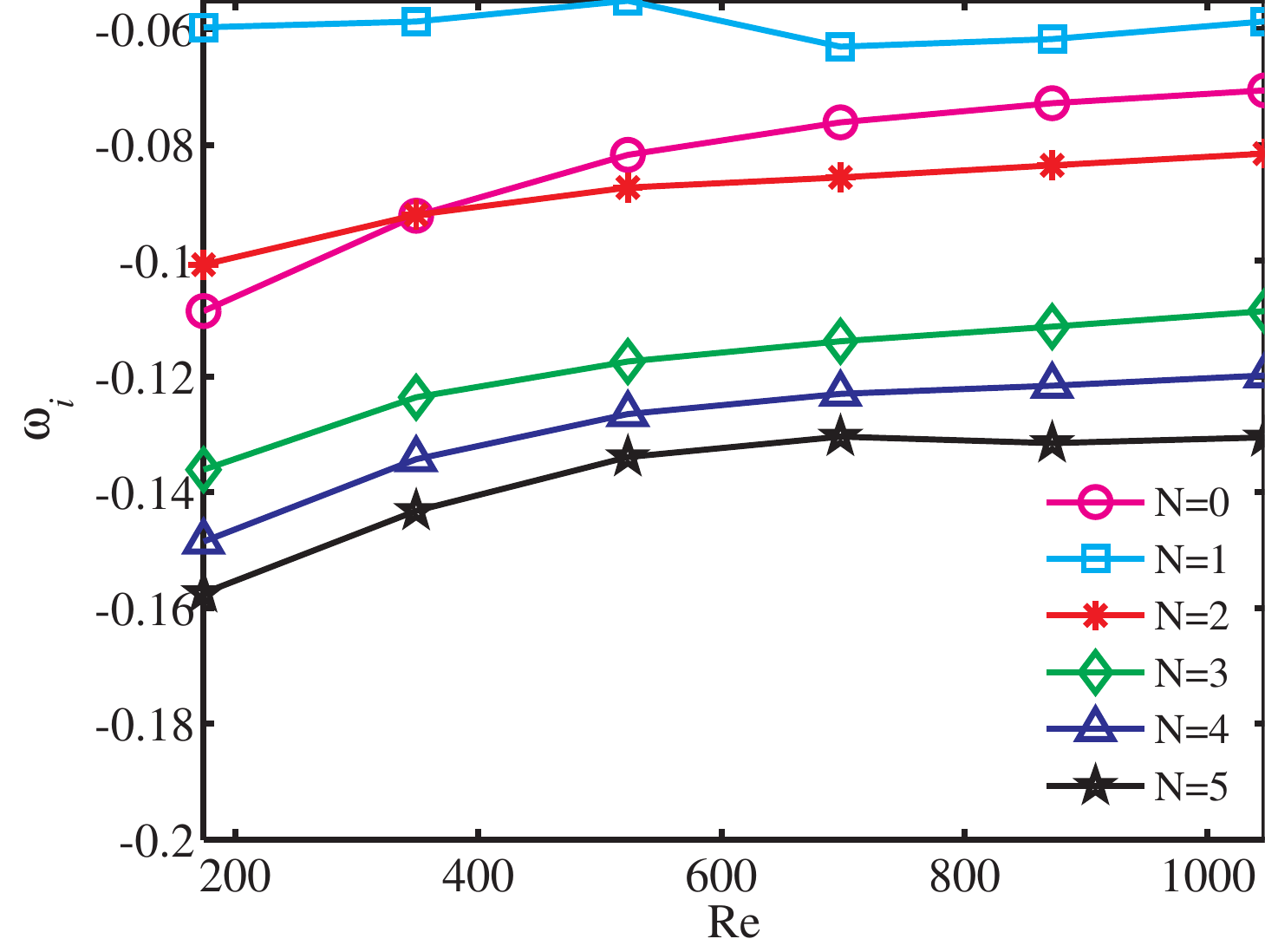} 
\end{center}
\begin{center}
\includegraphics[height=1.25in,width=1.65in, angle=0]
                {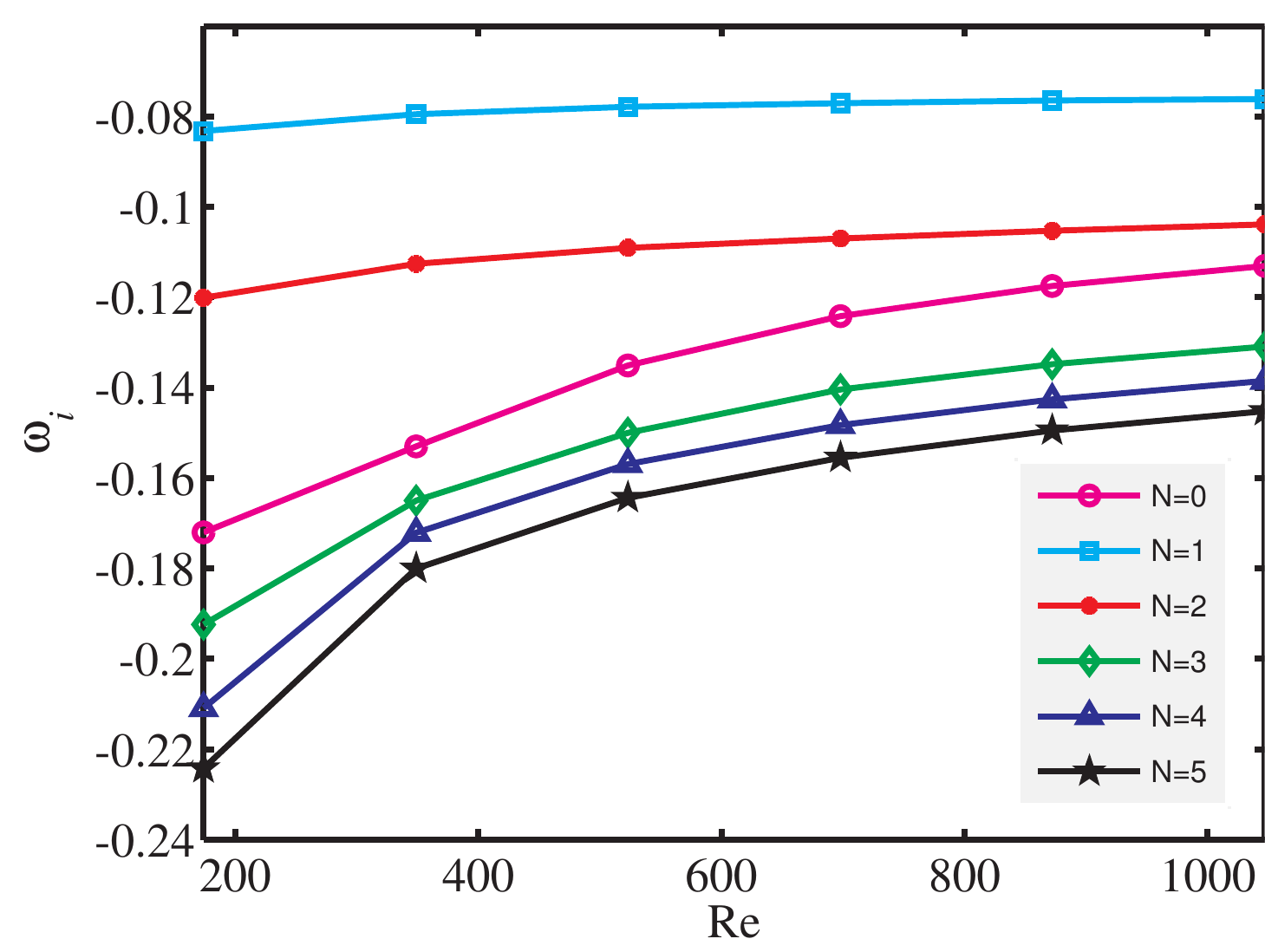}  
\caption{\label{tempgrowth} Variation in temporal growth rate $\omega_{i}$
with the Reynolds number for different semi-cone angles  
(a)$\alpha=2^0$ (b)$\alpha=4^0$ (c)$\alpha=6^0$.} 
\end{center}
\end{figure}

Figure \ref{tempgrowth} shows the temporal growth rate of the least stable 
eigenmodes for different Reynolds number and semi-cone angles $\alpha$.
The growth rate of eigenmodes increases with the  increase in Reynolds number for
all azimuthal wave-numbers(N) and semi-cone angles ($\alpha$).
For the Range of Reynolds number and semi-cone angles the least stable eigen
modes are negative, hence the Global modes are stable.
The Global modes with helical mode, $N=1$ are least stable for $\alpha=2^o$,
$\alpha=4^o$ and $\alpha=6^o$.
The damping rate of Global modes with $N=2$ is higher than $N=0$ for $\alpha=2^0$.
At smaller Re, $N=0$ is more stable than $N=2$, however at higher Re, 
$N=2$ is more stable then $N=0$ for semi-cone angle $\alpha=4^0$.
The helical modes $N=3$, 4 and 5  have larger damping rates then that of 
$N=0$, 1 and 2 for all Re and $\alpha$.
The Global modes are more stable at higher semi-cone  angle $\alpha$. 
The transverse curvature reduces in the stream-wise direction  with the 
increase in Reynolds number for same $\alpha$.
The favorable  pressure gradient is higher at higher semi-cone angle $\alpha$.
The Global modes are more stable at higher semi-cone angle, proves that favorable 
pressure gradient has damping effect on the Global modes.
\subsection{\label{sec:level1} Spatial amplification rate}
\begin{figure}
\begin{center}
\includegraphics[height=1.25in,width=1.65in, angle=0]
                {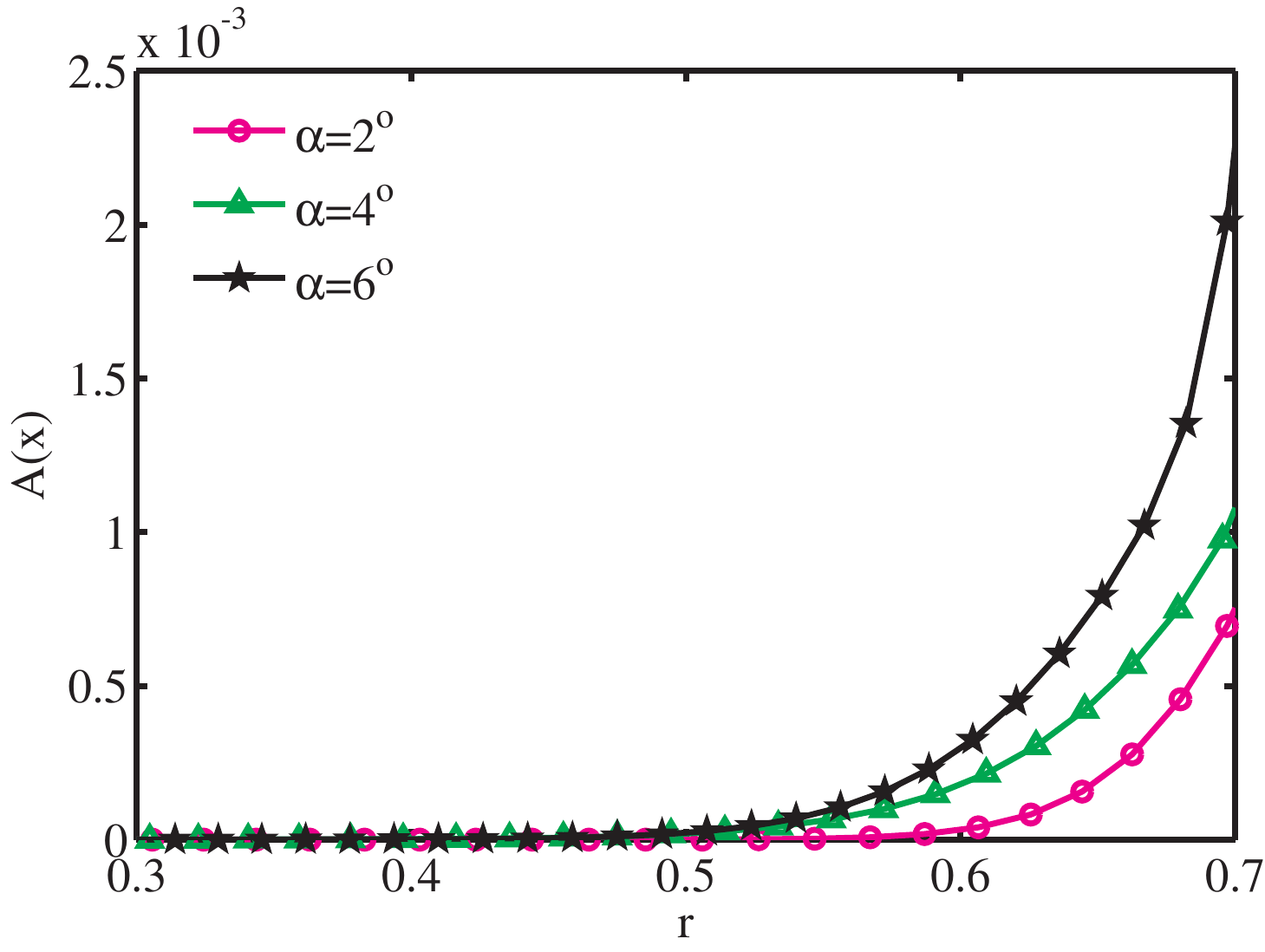}  
\includegraphics[height=1.25in,width=1.65in, angle=0]
                {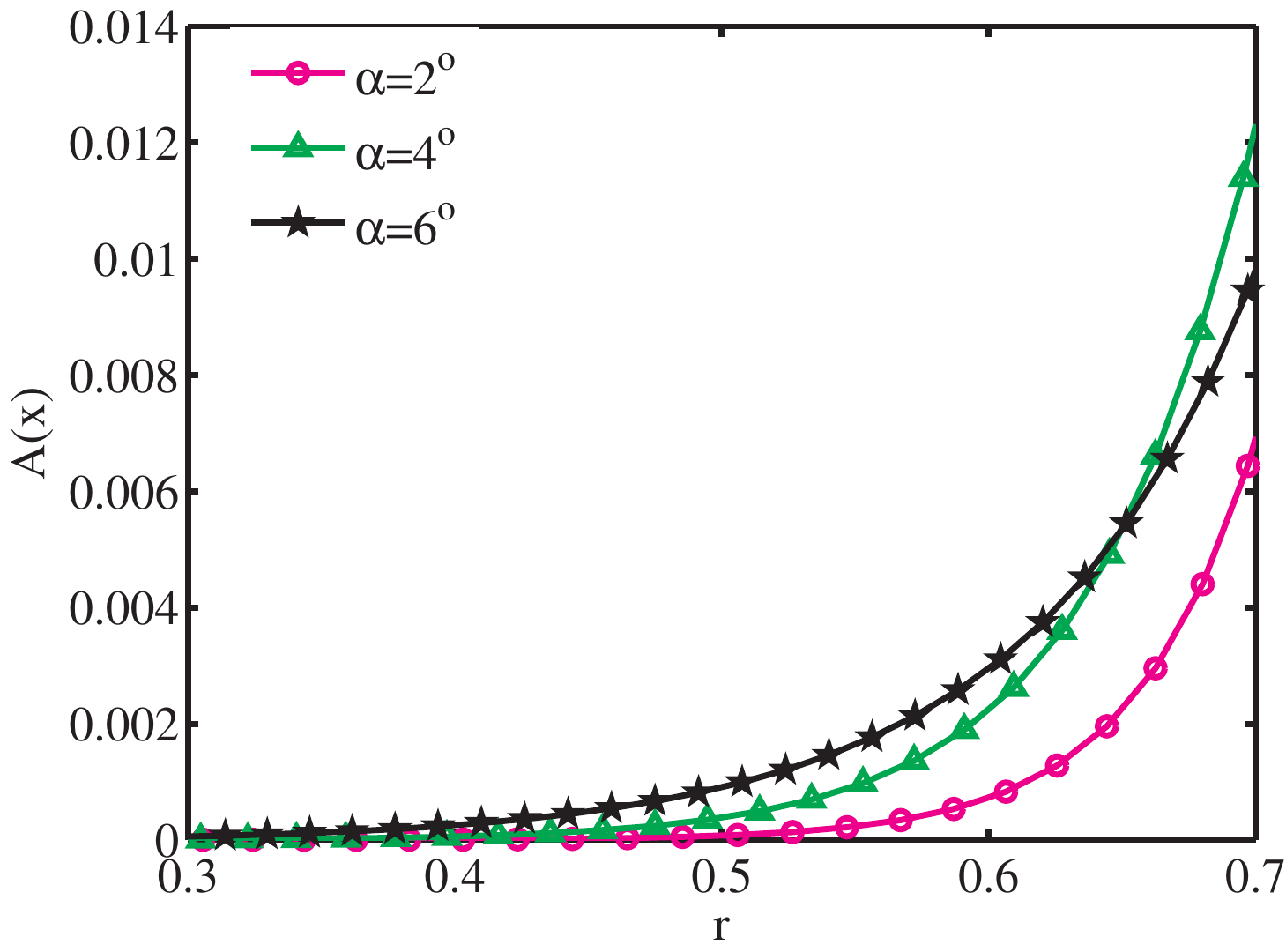} 
\end{center}
\begin{center}
\includegraphics[height=1.25in,width=1.65in, angle=0]
                {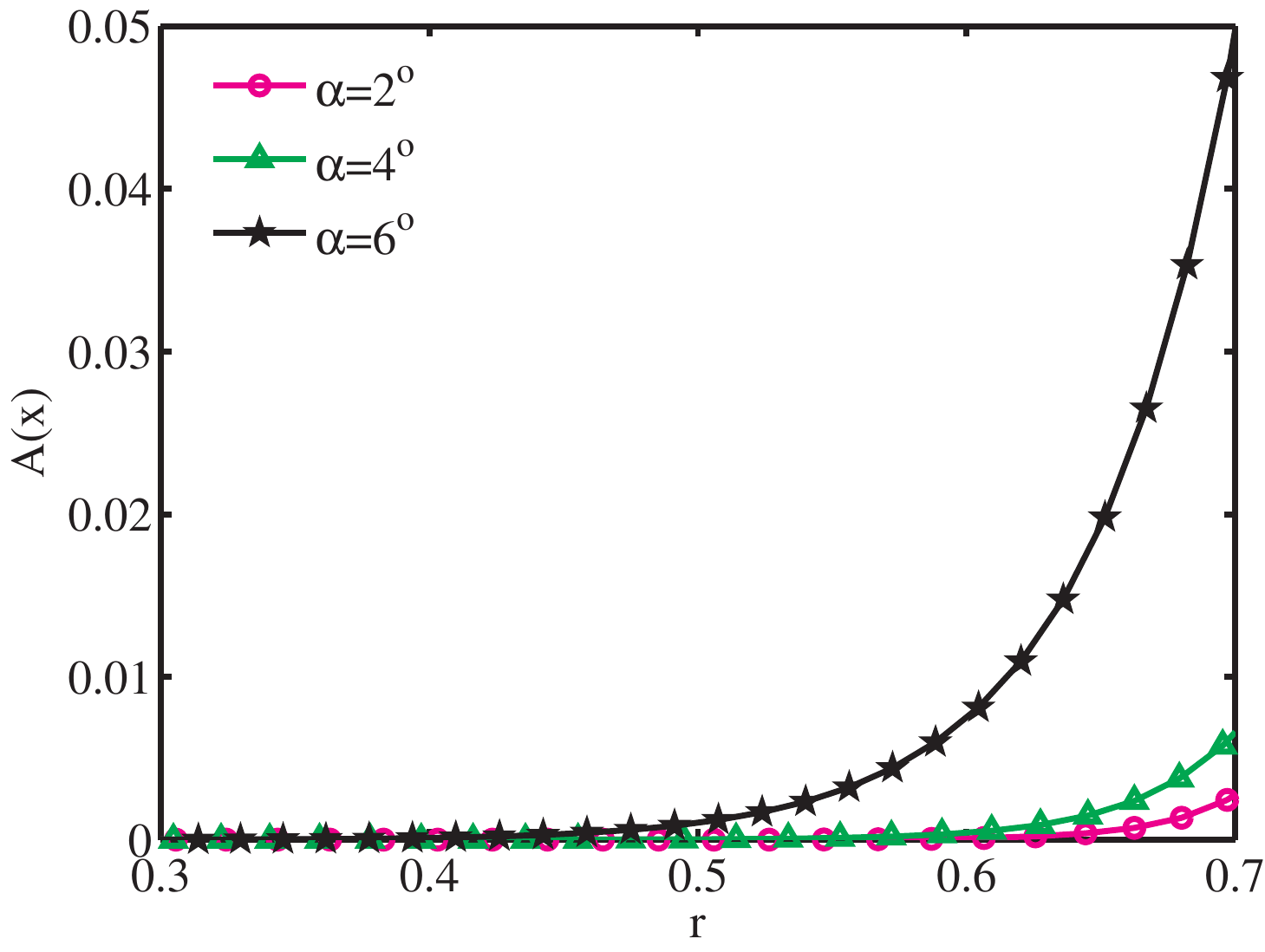}
\includegraphics[height=1.25in,width=1.65in, angle=0]
                {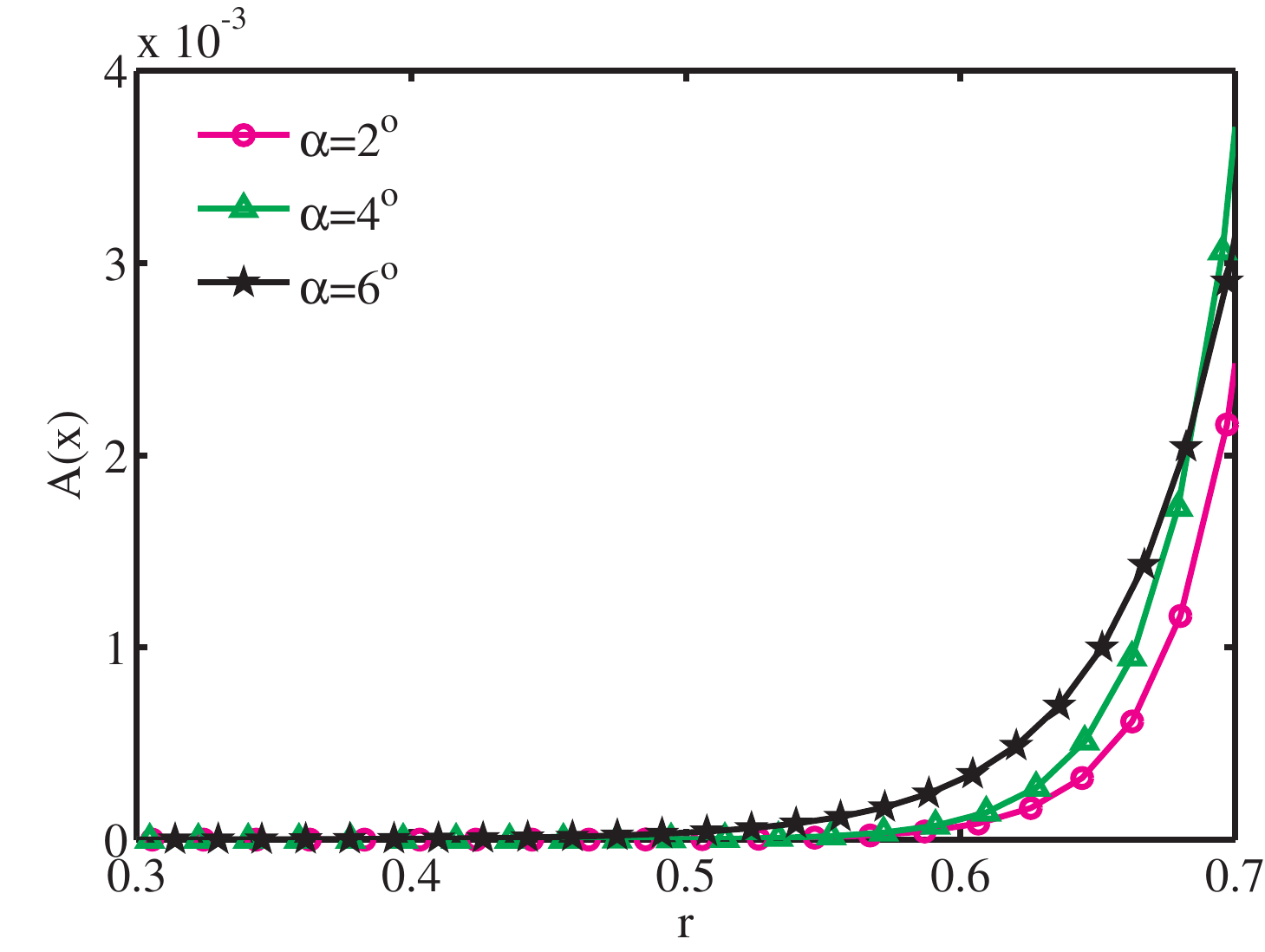}
\caption{\label{spgrowth} Variation of spatial growth rate $A_x$
in the stream-wise direction for different semi-cone angles $\alpha$ 
for Re=349. (a) $N=0$ (b) $N=1$ (c) $N=2$ and (d) $N=3$.} 
\end{center}
\end{figure}
The global temporal modes exhibits growth/decay in the stream-wise
direction while moving towards the downstream.
This growth/decay at different stream-wise station can be quantified 
by spatial amplification rate ($A_{x}$). 
The spatial amplification rate ($A_{x}$) shows the growth of
all the disturbances together in the stream-wise direction.
Figure \ref{spgrowth} shows the $A_x$ for different azimuthal 
wave-numbers(N) and semi-cone angles($\alpha$) for $Re=349$.
The spatial growth rate($A_x$) increases with the increase in
semi-cone angle($\alpha$) for all the azimuthal wave-numbers for 
given Reynolds number.We computed it for $Re=174$ to $Re=1047$, however
the result are presented for $Re=349$ only.
As semi-cone angle($\alpha$) increases the growth rate of the 
disturbances  increases in the stream-wise direction and makes 
the flow convectively unstable. 
Hence, at higher semi-cone angle flow  becomes convectively more unstable.
For $N=0$, 1 and 2 the disturbances exhibits higher spatial
growth rate at small Reynolds number, however for $N=3$, 4 and 5
spatial growth rate is high for higher Reynolds number.
\subsection{\label{sec:level1} Conclusions}
The linear global stability analysis of boundary layer forms on a circular i cone is performed.
The combined effect of transverse curvature and pressure gradient has been studied.
The Global temporal modes are computed for the axisymmetric boundary layer on 
a circular cone for the range of Reynolds number from 174 to 1046 with azimuthal
wave-number, N from 0 to 5 and semi-cone angle $\alpha$=$2^o$,$4^o$ \& $6^o$.
The largest imaginary part ($\omega_i$) of the computed Global modes are negative 
for all the Reynolds numbers(Re), azimuthal wave-numbers(N) and semi-cone angles($\alpha$). 
Hence, the Global modes are temporally stable.
The wave-like behaviour of the eigenmodes are found in the stream-wise direction.
The wave-length of the wavelet structure reduces with the increase in 
frequency($\omega_r$). 
The 2D spatial structure of the Global modes show that the size and magnitude of
the disturbance amplitudes increases while moving towards downstream, which proves that
the flow is convectively unstable.
The damping rate of the disturbances increases with the increase in semi-cone 
angle($\alpha$) from $2^o$ to $6^o$.
Thus, Global modes are more stable at the higher semi-cone angles($\alpha$).
At the same time with the increase in ($\alpha$) from $2^o$ to $6^o$, the spatial growth 
rate ($A_x$) also increases at a given Reynolds number, thus the flow becomes convectively 
more unstable at the higher semi-cone angles.
The azimuthal wave-numbers $N=3$, 4 and 5 have less temporal growth($\omega_i$) and 
spatial growth($A_x$) compare to $N=0$, 1 and 2.
The azimuthal wave-number $N=1$ is found to be least stable one for all the Re and $\alpha$. 
The increase in the semi-cone angle($\alpha$) develops favorable  pressure gradient and 
reduces transverse curvature effect.
Thus, the favorable  pressure gradient stabilizes the Global modes and effect of 
transverse curvature reduces.  
Thus, the role of favorable  pressure gradient is more effective than that of transverse 
curvature on flow stability.

\end{document}